\documentclass[aps, prd, reprint, longbibliography, nofootinbib,superscriptaddress, floatfix]{revtex4-2}
\pdfoutput=1

\usepackage{amsmath,amssymb,amsfonts}
\usepackage{mathrsfs}
\usepackage{bbm}
\usepackage{slashed}
\usepackage{graphicx}
\usepackage{comment}
\usepackage{verbatim}
\usepackage[T1]{fontenc} 
\usepackage[utf8]{inputenc}
\usepackage[colorlinks]{hyperref}
\usepackage{mathbbol}
\usepackage[dvipsnames]{xcolor}
\usepackage[normalem]{ulem}
\usepackage{svg}
\usepackage{flushend}
\usepackage{physics}
\usepackage{multirow}
\usepackage{graphicx}% Include figure files
\usepackage{dcolumn}% Align table columns on decimal point
\usepackage{bm}% bold math
\usepackage{tensor}
\usepackage{comment}
\usepackage[utf8]{inputenc} 
\usepackage{graphicx,color,overpic,mathtools}
\usepackage{amsthm,amsmath,amssymb,hyperref,mathrsfs}
\usepackage{braket,bm,bbm}
\usepackage{cancel}
\PassOptionsToPackage{normalem}{ulem}
\usepackage{ulem} 
\usepackage{physics}
\usepackage{float}
\usepackage[english]{babel}
\usepackage{graphicx}
\hypersetup{
    colorlinks=false,
    pdfborder={0 0 0},
}

\begin{document}

\title{Asymptotic Safety in Generalized Proca Theories}

\author{Lavinia Heisenberg}
\email[]{heisenberg@thphys.uni-heidelberg.de}
\affiliation{Institute for Theoretical Physics, University of Heidelberg, D-69120 Heidelberg, Germany}

\author{Alessia Platania}
\email[]{alessia.platania@nbi.ku.dk}
\affiliation{Institute of Physics, University of Graz,
Universit\"atsplatz 5, 8010 Graz, Austria}
\affiliation{Center of Gravity, Niels Bohr Institute, Blegdamsvej 17, DK-2100 Copenhagen \O, Denmark}
\affiliation{Niels Bohr International Academy, Niels Bohr Institute, Blegdamsvej 17, DK-2100 Copenhagen \O, Denmark}

\author{Sara Rufrano Aliberti}
\email[]{s.rufranoaliberti@ssmeridionale.it}
\affiliation{
Scuola Superiore Meridionale, Largo S. Marcellino 10, I-80138, Napoli, Italy
}
\affiliation{Istituto Nazionale di Fisica Nucleare (INFN), Sez. di Napoli, Complesso Universitario di Monte Sant’ Angelo, Edificio G, Via Cinthia, I-80126, Napoli, Italy}

\begin{abstract}
   Generalized Proca Theories are the most general higher-derivative extensions of a massive vector field  that retain second-order equations of motion. They are phenomenologically interesting as models of dynamical dark energy that, unlike scalar-tensor theories, can naturally accommodate cosmological anisotropies. A key open question is whether such theories can be fundamental. As a first step in this direction, we investigate whether they admit an ultraviolet completion within a quantum field theory framework, working with a truncation comprising up to four powers of the Proca field and up to two derivatives. We find a triplet of non-Gaussian ultraviolet fixed points, that lie very close to one another. Only one of them features a non-tachyonic Proca mass and could thus serve as a consistent ultraviolet completion for Generalized Proca Theories. We name it the Proca fixed point. We discuss its stability and contrast its features with those of the standard Reuter fixed point of the asymptotic safety scenario for quantum gravity and matter. In particular, we show that the Gaussian and Reuter fixed points lie on singular hypersurfaces of the flow of Generalized Proca Theories, yet can act as quasi-fixed points in certain regimes.
\end{abstract}

\maketitle

\section{Introduction}

In the quest to extend our understanding of fundamental interactions beyond the Standard Model (SM) and General Relativity (GR), Generalized Proca Theories (GPTs) have been proposed as consistent classical models where the Einstein-Hilbert (EH) dynamics is complemented by that of a massive vector field with higher-derivative self-interactions and non-minimal couplings to gravity. In these theories interactions are constructed to yield the most general higher-derivative dynamics compatible with second-order equation of motion~\cite{Heisenberg:2014rta}. This is done analogously to Horndeski theories in gravity, and ensures that Ostrogradsky instabilities are avoided, both at the classical~\cite{Heisenberg:2014rta,BeltranJimenez:2016rff,Heisenberg:2018vsk,BeltranJimenez:2019wrd} as well as quantum level~\cite{Heisenberg:2020jtr}.

Crucially, GPTs offer rich phenomenology across cosmology~\cite{DeFelice:2016yws,DeFelice:2016uil,deFelice:2017paw} and high-energy physics~\cite{DeFelice:2016cri,Heisenberg:2017hwb,Heisenberg:2017xda}, providing novel avenues for modeling dark energy, enabling homogeneous and isotropic cosmological solutions that drive cosmic acceleration without a cosmological constant~\cite{DeFelice:2016yws}, and probing deviations from GR in strong-field regimes. They may alleviate the $H_0$ tension, outperforming $\Lambda$CDM in data fitting \cite{Heisenberg:2020xak}.
Unlike scalar-tensor theories (such as Horndeski or beyond-Horndeski models), vector fields can also introduce anisotropies, breaking isotropy in cosmological models \cite{Heisenberg:2016wtr} and inflationary scenarios. In certain setups, these vector fields could also play a role for dark matter.

In the context of beyond-the-standard-model physics, GPTs can arise in theories with spontaneous symmetry breaking of gauge fields.
They can provide effective field theory descriptions of hidden sector gauge bosons (e.g., dark photons) with self-interactions beyond the minimal coupling to the SM.

Proca theories have hitherto been considered as classical phenomenological models; a key open question is then whether they can be embedded in a more fundamental framework. One-loop corrections to their classical Lagrangians reveal promising traits, including technical naturalness and quantum stability \cite{Heisenberg:2020jtr}. By imposing unitarity, locality, causality, and Lorentz invariance on an unknown ultraviolet (UV) completion, stringent constraints have been derived on the Wilson coefficients of Generalized Proca vector models \cite{deRham:2022sdl,deRham:2018qqo,Bonifacio:2016wcb}. Yet, it is unclear if such theories admit a UV completion within or beyond any known approach to quantum gravity and matter~\cite{Bambi:2023jiz,Basile:2024oms,Buoninfante:2024yth}, if their structure is preserved by the renormalization group (RG) flow, and if their resulting landscape is compatible with the aforementioned theoretical bounds. Addressing these questions is paramount to assessing the viability of GPTs. 

In this letter we 
make first steps in this endeavour, by initiating the study of UV completeness in GPTs. For concreteness, we restrict  ourselves to a quantum field theory (QFT) framework and we ask whether the RG flow of GPTs is UV complete, i.e., whether it possesses a  suitable UV fixed point akin to the ``Reuter fixed point'' of the asymptotic safety program~\cite{Knorr:2022dsx, Eichhorn:2022gku, Morris:2022btf, Martini:2022sll, Wetterich:2022ncl, Platania:2023srt, Saueressig:2023irs, Pawlowski:2023gym, Bonanno:2024xne}. Within this program, such a fixed point is known to exist in pure gravity~\cite{Reuter:2001ag, Lauscher:2002sq, Codello:2006in, Gies:2015tca, Gies:2016con, Hamada:2017rvn, Knorr:2017fus, Christiansen:2017bsy, Falls:2017lst, Falls:2020qhj, Knorr:2021lll, Knorr:2021slg, Kluth:2022vnq, Baldazzi:2023pep} and in gravity-matter systems whose field content and symmetries include those of the SM~\cite{Dona:2013qba,Christiansen:2017cxa,Pastor-Gutierrez:2022nki}. The approach extends beyond perturbative quantization by allowing gravitational couplings to flow towards an interacting regime at high energies, which is crucial to the non-perturbative renormalizability of the theory. Compared to the setup of the standard asymptotic safety program, GPTs are characterized by finitely many interaction terms and by a massive vector field which breaks the $U(1)$ symmetry. It is thus not obvious that a non-trivial fixed point in this case exists. In this work, we investigate for the first time whether Generalized Proca models admit such a UV completion. We focus on a truncated dynamics involving six interaction couplings and show that, within our approximations, a suitable fixed point exists. In particular, we find three candidate UV completions with very similar fixed-point values and properties, but only one of these ``twin'' fixed points---which we collectively call the \textit{Proca triplet}---has a non-tachyonic mass. This \textit{Proca fixed point} has five relevant directions, meaning that the landscape of effective field theories stemming from it~\cite{Basile:2021krr,Knorr:2024yiu} is parametrized by four dimensionless Wilson coefficients. Finding a fixed point is however not enough to claim that GPTs are UV complete: the Proca fixed point is largely not as stable and rapidly converging as the Reuter fixed point of the asymptotic safety program. Moreover, the number of relevant directions, which quantifies the predictivity of the theory, is larger than in the most conservative version of asymptotic safety. Finally, the presence of a physical longitudinal mode for the Proca field implies that the origin of theory space---the Gaussian fixed point (GFP), where the dimensionless Proca mass also vanishes---lies on a singular hypersurface of the Proca beta functions. The Reuter fixed point also lies on such a hypersurface, as it sits at the boundary of validity of GPTs, where the Proca mass vanishes and the $U(1)$ symmetry is fully restored. These features may have important theoretical and phenomenological consequences for the infrared (IR) limit of asymptotically safe GPTs.

\section{Setup and derivation of the flow}

The scope of this work is to determine whether GPTs can be UV completed within a QFT framework, i.e., whether their flow admits a suitable UV fixed point. In general, the Generalized Proca Lagrangian~\cite{Heisenberg:2014rta} contains up to five unknown functions of the Proca field $A_\mu A^\mu$ and up to four derivatives. 

In order to have a consistent truncation with all terms at a given order while keeping computations feasible, we consider all operators with up to two derivatives and to the fourth order in the Proca field. 
Concretely, we shall focus on the following truncation
\begin{widetext}
    \begin{align}
    &\mathcal{L}_{GPT} =\frac{1}{16\pi G}\left[2\Lambda - R\right] + \frac{Z_A}{4}F^{\mu\nu}F_{\mu\nu} + \frac{1}{2}G_2\,A^2 + G_3\,A^2\,\nabla_\mu A^\mu +G_4\,A^4 + G_{4,1}\left[A^2\,R + \left(\nabla_\mu A^\mu\right)^2 - \left(\nabla_\mu A_\nu\right)\left(\nabla^\nu A^\mu\right)\right]  \notag\\
    & + G_{4,2}\left[A^4\,R + 2\,A^2\left(\nabla_\mu A^\mu\right)^2-2A^2\left(\nabla_\mu A_\nu\right)\left(\nabla^\nu A^\mu\right) \right]+G_{4,4} A^2 \left( \nabla^\mu A^\nu \nabla_\mu A_\nu - \nabla_\mu A_\nu \nabla^\nu A^\mu \right)\, + G_{4,5}A^\mu\,A^\nu\,F_\mu^\alpha\,F_{\alpha\nu} \, ,
    \label{FullL}
\end{align}
\end{widetext}
where $\nabla_\mu$ denotes the covariant derivative associated with the metric $g_{\mu\nu}$, $A_\mu$ is the Proca field with mass $m=\sqrt{G_2}$, $A^2\equiv A_\mu A^\mu$, and the whole Lagrangian is written in Euclidean signature for later convenience. Notably, the mass term $G_2\,A_\mu A^\mu$ breaks the $U(1)$ gauge symmetry, allowing the longitudinal mode of the $A_\mu$ field to propagate.

We stress right away that the dynamics of GPTs is tuned to avoid ghost modes, but this comes at the price of making these theories very special: first, there is no symmetry principle protecting the structures in Eq.~\eqref{FullL}, and hence the RG flow is expected to generate more terms than those allowed by the ghost-free condition. Secondly, in the Generalized Proca Lagrangian~\eqref{FullL} different operators share the same interaction coupling. In a general RG treatment this is difficult to handle consistently, because in the absence of a symmetry principle, couplings associated with different operators run independently and typically in a different manner. At the same time, inserting a different coupling in front of each operator would break the structure of the Proca Lagrangian and would make our computation meaningless. As we shall detail later, to overcome this problem we will choose expansion points for the Proca and metric fields that avoid the ambiguity while also minimizing the complexity of our calculations.

We will study the properties of RG flow of GPTs using the Functional Renormalization Group (FRG)~\cite{Dupuis:2020fhh}. The latter is a tool to study the RG flow of a QFT by converting the functional integral over fluctuating modes into a functional integro-differential equation for the so-called Effective Average Action (EAA) $\Gamma_k$, with $k$ being an RG scale. This scale-dependent effective action results from the integration of modes with $p \gtrsim k^2$. Hence, $\Gamma_k$ interpolates between the bare action, corresponding to the limit $k\rightarrow\infty$, and the effective action, which is obtained in the limit $k\rightarrow 0$. Specifically, the EAA obeys a formally exact equation known as the Wetterich equation~\cite{Wetterich:1992yh}
\begin{equation}
    k\partial_k\Gamma_k =\frac{1}{2}\text{STr}\left[\left(\frac{\delta^2(\Gamma_k+\Delta S_k)}{\delta\hat{\phi}\delta\hat{\phi}}\right)^{-1}\,k\partial_k\left(\frac{\delta^2\Delta S_k}{\delta\hat{\phi}\delta\hat{\phi}}\right)\right],
    \label{Wetterich_eq}
\end{equation}
where $\hat{\phi}$ is the generic set of fluctuation fields considered, the supertrace STr includes a sum over internal indices as well as an integration over momenta, and $\Delta S_k$ is a scale-dependent mass term: it is quadratic in all fluctuation fields---structurally, $\Delta S_k=\int_x \hat{\phi}\mathcal{R}_k(\Delta/k^2)\hat{\phi}$, with $\Delta \equiv-\nabla^2$---and regulates the flow, implementing the Wilsonian shell-by-shell integration of fluctuating modes by suppressing those with $p^2\lesssim k^2$. To this end, the regulator $\mathcal{R}_k(\Delta/k^2)$ must appropriately suppress fast-fluctuating modes~\cite{Reichert:2020mja}.

The Wetterich equation is a formally exact equation, but it cannot be solved exactly; its resolutions requires to specify an ansatz for the EAA, in the form of an expansion scheme (e.g., a derivative or vertex expansion) and a truncation order to start with, which can then be systematically improved. Within such truncated flows, both the interaction couplings and the unphysical RG scale $k$ can take any values. Hence, the flow is non-perturbative in the couplings, but perturbative in the parameter(s) governing the expansion of the effective action (e.g., physical momenta in the case of a derivative expansion, and field strengths in the case of a vertex expansion---everything measured in appropriate units). Keeping these caveats in mind, in the following we scratch the key steps in the derivation of the FRG flow for the ansatz~\eqref{FullL}. 

To write down the Wetterich equation for gauge theories, and in particular for gravity, it is useful to expand each field about a fixed yet arbitrary background. This step is at the core of the background field method and is important to define the gauge-fixing and Faddeev-Popov actions, as well as covariant derivatives and the RG scale in the presence of dynamical and fluctuating gravity. In our case, we need to introduce the background fields $\{\bar{g}_{\mu\nu},\bar{A}_\mu\}$ and fluctuating fields $\{h_{\mu\nu},\hat{A}_{\mu}\}$ for  the metric and Proca fields. In particular, we shall use a linear split of metric and vectorial fluctuations, namely, we shall define $g_{\mu\nu}={\bar{g}}_{\mu\nu}+h_{\mu\nu}$ and ${A}_{\mu}=\bar{A}_{\mu}+\hat{A}_{\mu}$. With this, the left-hand-side of the Wetterich equation is directly evaluated on the arbitrary background fields $\{{\bar{g}}_{\mu\nu},\bar{A}_{\mu}\}$, whereas its right-hand-side is evaluated on them after computing the second variation of $\Gamma_k$. In particular, the full action---including the gravitational gauge-fixing and Faddeev-Popov actions---reads
\begin{equation}\label{eq:full-action}
    \begin{aligned}
        \Gamma_k &= \int d^4x\sqrt{g}\left[\mathcal{L}_{GPT} +\frac{1}{2\alpha} \bar{g}^{\mu\nu} F_\mu F_\nu   -\frac{\bar{C}_\mu \left(\Delta_{gh}\right)_\nu^{\mu} C^\nu}{16\pi G(k)} \right], 
    \end{aligned}
\end{equation}
where $\mathcal{L}_{GPT}=\mathcal{L}_{GPT}(h,\hat{A}; \bar{g}, \bar{A})$ is the Lagrangian in Eq.~\eqref{FullL}, with all interaction couplings replaced by their running counterparts, $\{\Lambda,G,G_i\} \to \{\Lambda(k),G(k),G_i(k)\}$. Moreover, $\bar{\nabla}_\mu$ is the covariant derivative on the background metric ${\bar{g}}_{\mu\nu}$, the tensor
\begin{equation}
F_{\mu}\left(h;\bar{g}\right)=\frac{1}{16\pi G(k)} \left(\bar{\nabla}_\nu \,h^\nu_\mu - \frac{1+\beta}{4}\bar{\nabla}_\mu h \right)
\end{equation} 
implements the gauge condition $F_\mu\left(h;\bar{g}\right)=0$ for the gravitational sector, with the constants $\alpha$ and $\beta$ being gauge fixing parameters. Finally, $\bar{C}_\mu$ and $C_\mu$ are the gravitational Faddeev-Popov ghost fields, and 
\begin{equation}
\left(\Delta_{gh}\right)_\nu^{\mu} = \left(\bar{\nabla}^2 \delta^\mu_\nu + \frac{1-\beta}{2} \bar{\nabla}_\nu \bar{\nabla}^\mu +\bar{R}_\nu^\mu\right)
\end{equation}
is the Faddeev-Popov operator. It is important to notice that since the $U(1)$ gauge symmetry is explicitly broken in GPTs, no gauge fixing is necessary in the Proca sector. This makes the longitudinal mode of the $A_\mu$ field \textit{physical}. Nonetheless, the corresponding propagator is ``static'', since no kinetic term turns out to be associated with it, akin to the case of auxiliary fields. These aspects make the flow of GPTs fundamentally different than that of a general Einstein-Maxwell Lagrangian with standard symmetries and all terms allowed by them. Moreover, these features generate an ambiguity on whether the longitudinal mode should be regularized in the spirit of the RG: physical modes should in principle be regularized, but the absence of a kinetic term makes the Wilsonian shell-by-shell integration of modes ineffective. In this work we shall regularize the longitudinal mode, based on three lines of reasoning: \textit{(i)} to treat all physical modes on the same footing, \textit{(ii)} to avoid spurious scheme dependencies~\cite{Litim:1998qi,Litim:1998nf,Braun:2010tt}, and \textit{(iii)} to avoid highly non-linear beta functions and related conditional expressions, which are found by explicit computations.

Since the regulator $\Delta S_k$ must be quadratic in all fluctuation fields, its most general form is
\begin{align}\label{eq:regulator}
    \Delta S_k &= \frac{1}{2}\int d^4x \sqrt{\bar{g}}\,h_{\mu\nu}\,\mathcal{R}_g^{\mu\nu\rho\sigma}\,h_{\rho\sigma}+\int d^4x \sqrt{\bar{g}}\,\bar{C}_\mu \mathcal{R}^{\,\,\,\,\,\,\mu}_{gh\,\,\nu}\,C^\nu \notag\\
    &+ \frac{1}{2}\int d^4x \sqrt{\bar{g}}\,\hat{A}_\mu \,\mathcal{R}^{\mu\nu}_A\hat{A}_\nu,
\end{align}
where we have introduced the $k$-dependent regulators whose tensorial structure reflects that of the corresponding two-point functions and whose scalar part $R_k(\bar{\Delta}/k^2)$---known as ``shape function''---has suitable suppression properties that guarantee UV and IR regularity of the flow~\cite{Percacci:2017fkn,Reuter:2019byg,Reichert:2020mja,Basile:2024oms}. In our work we shall employ a ``type I'' regularization scheme, in which each regulator depends only on the background Laplace operator $\bar{\Delta}$. Moreover, in our implementation of the flow we shall use the Limit shape function, $R_k(p^2/k^2)=(k^2-p^2)\,\theta(k^2-p^2)$.

At this point we can use the background field method and background independence to our advantage to simplify computations, while keeping a consistent truncation retaining all operators according to the chosen scheme, Eq.~\eqref{FullL}. To this end, we shall choose a constant background Proca field $\bar{A}_\mu$ and an Einstein manifold for the gravitational background $\bar{g}_{\mu\nu}$. This reduces the number of couplings from ten to six, and removes the ambiguity we mentioned earlier, of introducing additional couplings of the $G_{4,i}$-type for each operator in the squared brackets in Eq.~\eqref{FullL}. In this way we preserve the structure of Proca theories while minimizing the number of couplings in a consistent truncation. Specifically, with our choice of background, we will be able to resolve the running couplings $\{G,\Lambda,G_2,G_{4,1},G_4,G_{4,2}\}$. Note that, while the pure-derivative couplings like $G_3$ and $G_{4,5}$ cannot be resolved within our setup, the operators containing derivatives of the Proca fields and multiplying the couplings $G_{4,1}$ and $G_{4,2}$ still influence the flow. This is because they enter the second variation of the EAA and hence contribute to the supertrace.
To further simplify computations, we can reduce the number of off-diagonal terms in the Hessian $\Gamma_k^{(2)}$ by using the Feynman gauge $\alpha=\beta=1$ for the gravitational field. This simplifies the inversion of the regularized two-point function in  Eq.~\eqref{Wetterich_eq}. 

With these details in mind, we can now replace each ingredient---Eqs.~\eqref{eq:full-action}-\eqref{eq:regulator}---in the Wetterich equation~\eqref{Wetterich_eq}, where now
$\hat{\phi}=\left\{h,\hat{A},\bar{C},C,\bar{C}_A,C_A\right\}$. At this point, the calculation of the left-hand-side of the flow equation~\eqref{Wetterich_eq} for our ansatz~\eqref{FullL} is straightforward: introducing the dimensionless counterparts to the running couplings of the system,
\begin{align}
    \lambda(k)&\equiv \Lambda(k)k^{-2},\qquad &g(k)&\equiv G(k)k^{2}, \notag \\
    g_2(k)&\equiv G_2(k)k^2,\qquad &g_{4,1}(k)&\equiv G_{4,1}(k),\notag \\ 
    g_4(k)&\equiv G_4(k),\qquad &g_{4,2}(k) &\equiv G_{4,2}(k) k^{-2},
\end{align}
the beta functions $\beta_{g_i}\equiv k\partial_k g_i$ appear as the coefficients of the interaction operators in the variation $k\partial_k\Gamma_k$. 
Spelling out the right-hand-side of Eq.~\eqref{Wetterich_eq} involves instead a few steps, which can sometimes be non-trivial~\cite{Basile:2024oms}: the calculation of the regularized two-point function and its inversion, the contraction with the variation of the regulator, and the computation of the supertrace. As anticipated, the first step is made feasible by our setup, which uses simple background fields and a harmonic gauge fixing. However, not all open derivatives cancel due to the longitudinal mode of the Proca field being physical and hence ungauged in GPTs. The inversion of the regularized two-point function cannot be done exactly though, and some manipulations are necessary, which include a Parker-Fulling expansion of the propagator (see App.~\ref{Appendix:PF-exp}). This step of the computation is the most involved, as the inversion of the regularized two-point function produces terms involving different orderings of curvature invariants, open derivatives and functions of the Laplacian operator, so that a direct application of the heat kernel techniques to the supertrace argument is not possible. To overcome this issue, one first needs to systematically re-arrange these operators~\cite{Knorr:2021slg} by means of general commutation relations between open derivatives and arbitrary functions of the Laplacian. These have been derived in~\cite{Knorr:2021slg}, and we review them in App.~\ref{Appendix:Commutator_rules}, adapting them to our analysis.

Once both sides of the Wetterich equation for $\bar{\Gamma}_k\left(\bar{g},\bar{A}\right)$ are evaluated, comparing the same functional integrals appearing on both left- and right- hand sides allows one to extract the beta functions for each dimensionless coupling. We computed them using standard \texttt{xact} packages~\cite{Martin-Garcia:2007bqa} and can be found in the attached mathematica notebook.

\section{Asymptotic safety in Generalized Proca Theories}\label{Sec_Asymptotic_safety_in_Generalized_Proca_Theories}

The beta functions encode the behavior of the running couplings at all energy scales, and their zeros define the possible IR and UV limits of the theory. In order to determine whether GPTs are UV complete, as a first step we need to solve the fixed-point equations
\begin{equation}
    \beta_i(g_j^\ast)=0\,.
\end{equation}
Here $\beta_i$ are the beta functions associated with the dimensionless running couplings $g_j\equiv\{\lambda, g, g_2, g_{4,1}, g_4, g_{4,2}\}$ and $g_j^\ast$ denotes their fixed-point values.
This system of equations can generally admit two types of solutions: a GFP, which is the origin of the theory space and is expected on general grounds, and potentially one or more non-Gaussian fixed points (NGFPs), where the dimensionless couplings can take any real non-zero value. The former fixed point describes a free theory, while the latter correspond to interacting theories. Assuming one such fixed points exist, whether it can serve as a predictive UV completion for the theory depends on its critical exponents $\theta_i$. These are defined as the opposite of the eigenvalues of the stability matrix $S_{ij}\equiv \partial_{g_j}\beta_i|_{g^*}$, and yield the stability properties of a fixed point: a positive $\theta_i$ indicates that the associated eingendirection corresponds to an IR-relevant direction, i.e., initializing the flow close to the fixed point, and on this eigendirection, the corresponding RG trajectory attains the fixed point in the UV limit $k\to\infty$. In the opposite limit, $k\to0$, IR-relevant directions correspond to parameters to be measured. Vice versa, IR-irrelevant directions correspond to predictions in the IR. Hence, a fixed point provides a predictive UV completion for the theory if \textit{(i)} it has at least one relevant direction, so that at least one RG trajectory can be UV-completed by this fixed point, and \textit{(ii)}, the number of relevant direction is finite and ideally small, so that only a few experiments have to be performed in the IR to completely fix the theory.

The non-linearity of the system of fixed-point equations produces a number of solutions, but it is expected that not all of them correspond to true UV completions of the theory: when truncating the dynamics, the fixed-point equations can have multiple solutions, but most of them are typically truncation artifacts that disappear as the truncation order is increased~\cite{Falls:2017lst,Falls:2016wsa,Falls:2014tra,Falls:2013bv,deBrito:2023myf}. For instance, in the context of asymptotically safe gravity, increasing the truncation order shows how spurious fixed points disappear while the Reuter fixed point and its critical exponents remain stable~\cite{Reuter:2001ag, Lauscher:2002sq, Codello:2006in, Dona:2013qba,Gies:2015tca, Gies:2016con, Biemans:2016rvp,Christiansen:2017cxa, Hamada:2017rvn, Knorr:2017fus, Christiansen:2017bsy, Falls:2017lst, Falls:2020qhj, Knorr:2021lll, Knorr:2021slg, Kluth:2022vnq, Pastor-Gutierrez:2022nki, Baldazzi:2023pep}. 
\begin{figure}[t!] 
\includegraphics[width=0.43\textwidth]{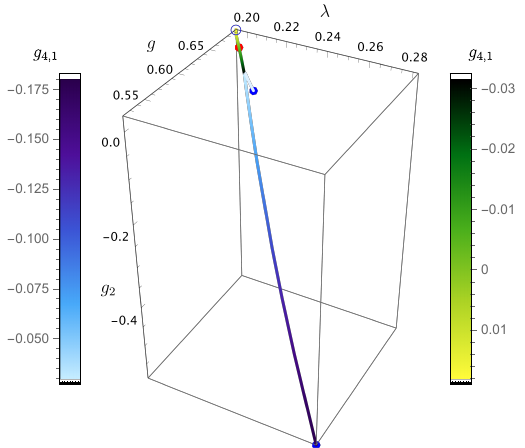}
    \caption{The figure shows how the fixed-point structure of GPTs changes upon introducing the first non-trivial interaction term. Turning $g_{4,1}$ on causes the Reuter fixed point (red dot) to split into three fixed points (blue dots), one of which is a pair of complex-conjugate fixed points (empty blue dot) at the full-$A^2$ order. This is revealed by tracking the evolution of the triplet as $g_{4,1}$ is gradually switched on. The connecting lines follow the zeros of the first three beta functions until the fourth one, $\beta_{g_{4,1}}$, also vanishes for specific values of $g_{4,1}^\ast$. For $g_{4,1}=0$, the Proca and gravitational fields interact minimally and the relevant fixed point is the Reuter one (red dot). Increasing $g_{4,1}$ to positive values, all four beta functions vanish at $\rm{Re}(g_{4,1}) \simeq 0.032$, yielding a complex-conjugate pair (empty blue dot). For negative $g_{4,1}$, the partial fixed point eventually splits into two real fixed points, reached at $g_{4,1}^\ast \simeq -0.072$ and $g_{4,1}^\ast \simeq -0.58$ (blue dots). No fixed-point collision occurs, since $\{\beta_\lambda,\beta_g\}$ depend linearly on~$g_{4,1}$, while~$\beta_{g_2}$ is cubic in it.}
    \label{fig:quasi}
\end{figure}
In order to determine whether Proca theories admit a UV completion that is not a truncation artifact, and to compare it to the Reuter fixed point, we performed our analysis starting from the EH truncation and reaching the maximal truncation~\eqref{FullL} by adding one coupling at a time. In this way, we enlarge the truncation in a step-by-step fashion. This strategy allows us to identify candidates for the UV completion of GPTs: these are fixed points whose coordinates vary continuously when adding higher-order terms, and whose properties are stable across different truncations. 

The candidate fixed points were analyzed and tracked across different truncations not only by comparing the values of the dimensionless couplings, but also by examining the associated critical exponents. This is key because tracking the continuation of the fixed point(s) beyond the EH truncation in GPTs is nontrivial: already upon adding a single Proca coupling to the EH Lagrangian, multiple fixed points appear. To track the evolution of these fixed points across successive truncations and identify the spurious ones, one must determine which fixed point at a given order corresponds to which one at the previous order. However, as we shall see, the fixed points of GPTs lie close to one another in theory space, making such identification ambiguous when solely based on the values of the fixed-point coordinates. The critical exponents, on the other hand, differ more significantly and thus provide a reliable criterion to distinguish between nearby fixed points and follow the evolution of each candidate UV completion as the truncation order is increased. Hence, to trace the evolution of the individual fixed points in GPTs and determine their stability properties it is particularly important to study, along the fixed-point values, also their critical exponents.

\begin{figure}[t!]
\includegraphics[width=0.475\textwidth]{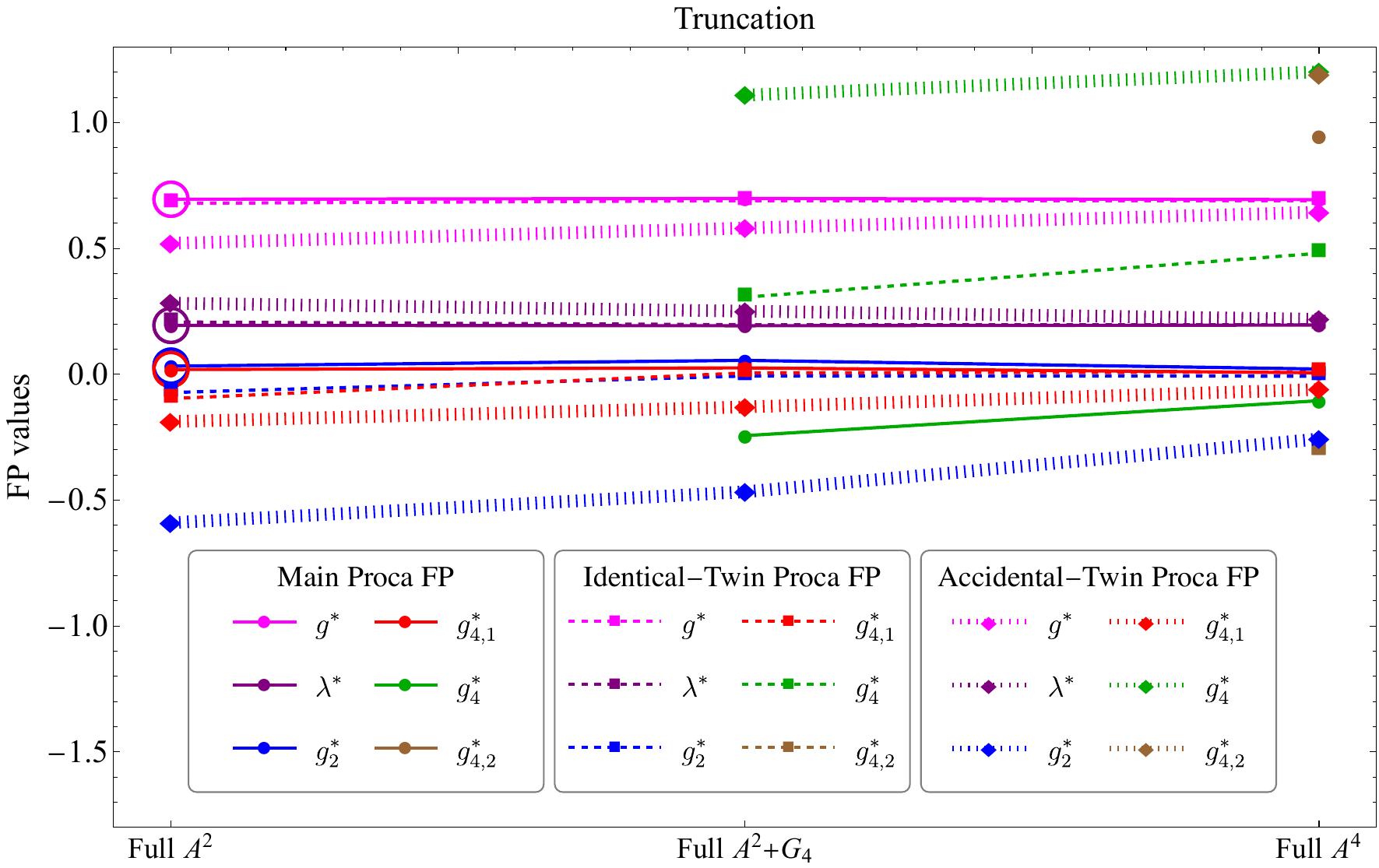}
    \caption{Evolution of the fixed-point values for the dimensionless couplings of the main (circles, connected by solid lines), identical-twin (squares, connected by dashed lines), and accidental-twin (rhombuses, connected by vertically dashed lines) Proca fixed points as a function of the truncation order. At each step, we add one coupling to control the evolution and stability of the Proca fixed point and its twins; we start from the full-$A^2$ truncation, since this is the lowest order truncation where the Proca fixed point and its twins appear. Filled circles, squares, and rhombuses correspond to real fixed points, whereas empty ones denote pairs of complex-conjugate fixed points. The main Proca fixed point emerges from a pair of complex-conjugate fixed points, but it is the only one with a non-tachyonic mass.} 
    \label{fig:FPvsTruncation}
\end{figure}

\begin{figure}
    \centering
    \includegraphics[width=1
    \linewidth]{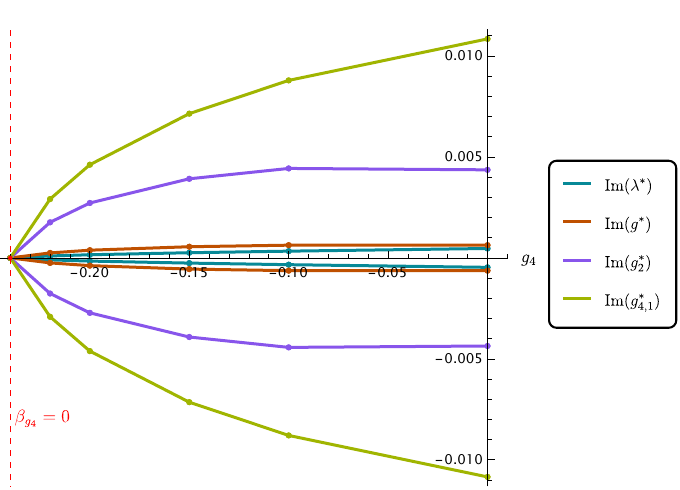}
    \caption{Merger of two complex-conjugate fixed points into a real one as the coupling $g_{4}$ is slowly activated. Data are produced by varying $g_{4}$ from zero to the fixed-point value $g_{4}^\ast\approx -0.24$, and solving the system $\beta_\lambda=\beta_g=\beta_{g_2}=\beta_{g_{4,1}}=0$ for each intermediate $g_4$. The colorful lines show how the imaginary parts of the fixed-point values of the other couplings diminish to eventually vanish when the collision happen. At the same point, also the beta function $\beta_{g_4}$ vanishes. Through this mechanism, the main Proca fixed point turns from a complex-conjugate pair into a real fixed point upon inclusion of the interaction coupling $g_4$.}
    \label{fig:merger}
\end{figure}

The quantitative results of our analysis are listed in the tables in App.~\ref{appendix:Tables}. For each truncation we solved the fixed-point equations looking at the continuation of the fixed points found in the previous step. In the first step, i.e., in the EH truncation, there is only one fixed point: the Reuter fixed point of the traditional asymptotic safety program. Including the Proca-mass term in the system does not change this picture: there is a continuation of the Reuter fixed point, characterized by a vanishing Proca mass, alongside with two additional NGFPs which can be considered spurious, as they disappear in higher-order truncations\footnote{More precisely, one can explicitly verify that they become and remain complex in all higher-order truncations we considered.}. The fixed point structure undergoes a qualitative change when the first non-minimal interaction between the Proca and gravitational field is included. As shown in Fig.~\ref{fig:quasi}, by slowly activating the~$g_{4,1}$ coupling, the Reuter fixed point splits into three new NGFPs that lie close to one another; we will collectively refer to them as the \textit{Proca triplet}. The evolution of the Proca triplet across different truncations, starting from the full-$A^2$ one, is summarized in Fig.~\ref{fig:FPvsTruncation} (see also App.~\ref{appendix:Tables}). 
Within the Proca triplet, only one fixed point has a non-tachyonic mass,  $m^2(k)\propto g_2^*k^2>0$, and can therefore be regarded as a viable candidate for a consistent UV completion of GPTs. In the following, we refer to it as the ``main'' Proca fixed point, or simply the \textit{Proca fixed point}, while the other two in the triplet, which lie very close to it, are denoted as ``twin'' Proca fixed points. The Proca fixed point first appears at the full-$A^2$ order as a complex-conjugate pair\footnote{We stress that this pair does not arise from the collision of two fixed points from a previous step, and this is due to the structure of the beta functions of $\{\lambda,g,g_2\}$, them being  linear and cubic in $g_{4,1}$ (cf. Fig.~\ref{fig:quasi}).}---with a small but non-vanishing imaginary part---but becomes and remains real at higher-order truncations. 
The merger of the complex-conjugate pair into a single real fixed point is illustrated in Fig.~\ref{fig:merger}. The Proca mass associated with this fixed point is always real and positive. Nevertheless, its emergence from a complex-conjugate pair may signal slow convergence. This interpretation is supported by the evolution of its critical exponents, which we report in detail in App.~\ref{appendix:Tables}, cf.~Fig.~\ref{fig:ReCritExpvsTruncation}. Whenever present, the main Proca fixed point is given in the first row of the tables in App.~\ref{appendix:Tables}. 

As already remarked, all fixed points in the Proca triplet originate from the same fixed point (cf. Fig.~\ref{fig:quasi}). The two ``twin'' Proca fixed points are however tachyonic. One of them, the \textit{identical-twin Proca fixed point} (second row in the tables in App.~\ref{appendix:Tables}), has coordinates and critical exponents nearly identical to those of the (main) Proca fixed point in all truncations, with the sole qualitative difference being the sign of the Proca mass squared.  In contrast, the \textit{accidental-twin Proca fixed point} (third row in the tables in App.~\ref{appendix:Tables}) emerges further away at the full-$A^2$ order, but approaches the main Proca fixed point as the truncation order is increased. Yet, its critical exponents are noticeably different than those of the main and identical-twin Proca fixed points. In particular, their magnitude signal that this fixed point is highly non-perturbative.

Along with the fixed point coordinates, we analyzed and tracked the evolution of the critical exponents of the Proca triplet. The specific values are reported in the tables in App.~\ref{appendix:Tables} and summarized in Fig.~\ref{fig:ReCritExpvsTruncation}. The real part of the critical exponents associated with the first two eigendirections are relatively stable for all fixed points, i.e., their magnitude does not vary significantly across truncations. The other critical exponents vary more substantially. The stability of their sign across different truncations is particularly important, as it determines whether the corresponding eigendirection is relevant or irrelevant. In our case, some of the signs change across truncations. In the last step, the main and accidental-twin Proca fixed points feature five relevant directions, whereas the identical-twin fixed point has four, making it the most predictive of the Proca triplet. Its viability is however compromised by the tachyonic Proca mass at the fixed point\footnote{In principle, resumming quantum effects could generate additional minima in the potential, with respect to which the Proca field would not be tachyonic~\cite{Platania:2022gtt}. However, this possibility is speculative in the absence of explicit computations, and we therefore disregard it.}.
As for the imaginary parts of the critical exponents, three of them are stable, whereas the two associated vary more significantly.

Overall, due to the slow convergence of (some of the) critical exponents of the Proca fixed point, a definitive assessment of its existence and viability as a UV completion for GPTs requires going beyond the present approximation, which includes operators up to second order in derivatives and fourth order in the vector field. Two outcomes might be possible: either the main Proca fixed point exists but converges slowly, or GPTs do not admit a UV completion within a QFT framework. In particular, the slow convergence of the Proca fixed point and its twins may be due to the fine-tuned dynamics of GPTs, which is likely not preserved by the RG flow in the absence of a symmetry protecting the structure of its interactions.

\begin{figure}[t!]
\includegraphics[width=0.45\textwidth]{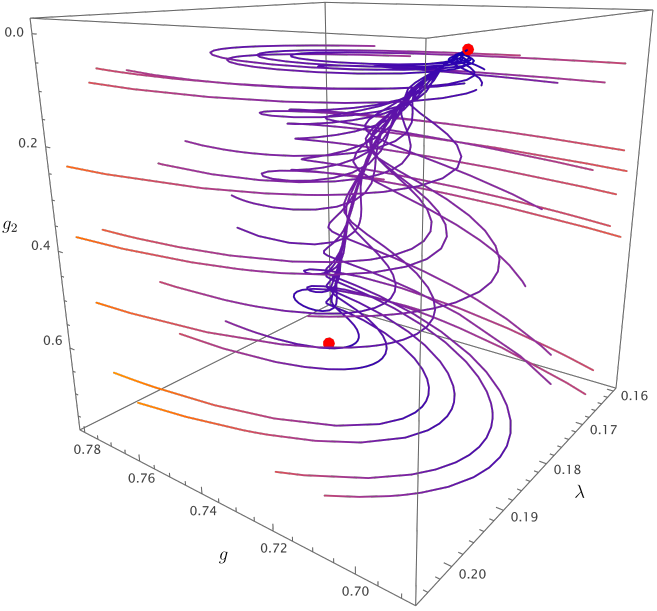}
    \caption{Projection of the full flow onto the $\{\lambda,g,g_2\}$ sub-theory space. The projected main and identical-twin Proca fixed points (red dots) have complex-conjugate critical exponents, which cause the characteristic spiraling behaviors of close-by RG trajectories. The two NGFPs are connected by a separatrix line.} 
    \label{fig:3dstep2}
\end{figure}

\begin{figure*}[t!]
\includegraphics[width=0.32\textwidth]{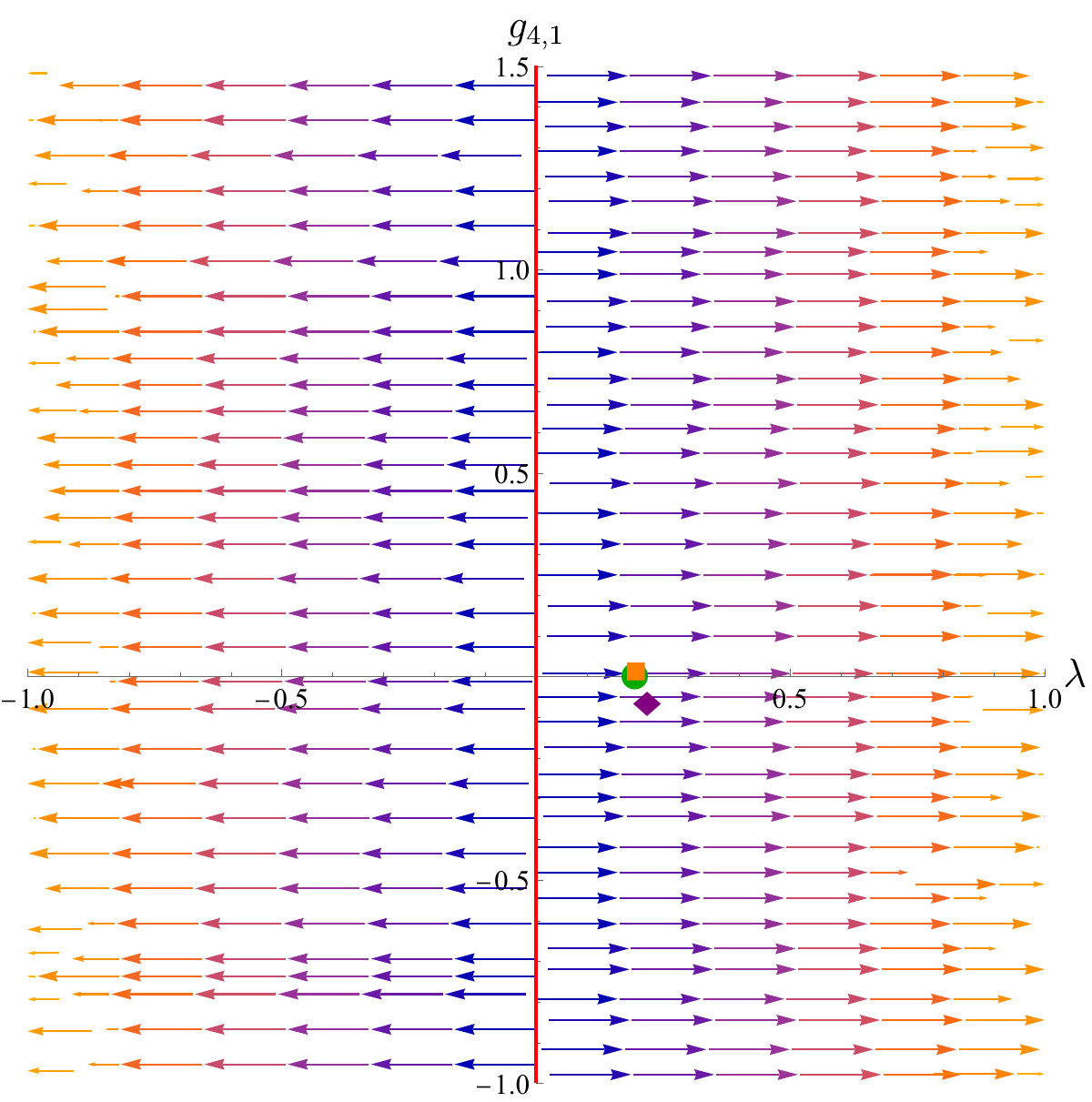}\hfill\includegraphics[width=0.32\textwidth]{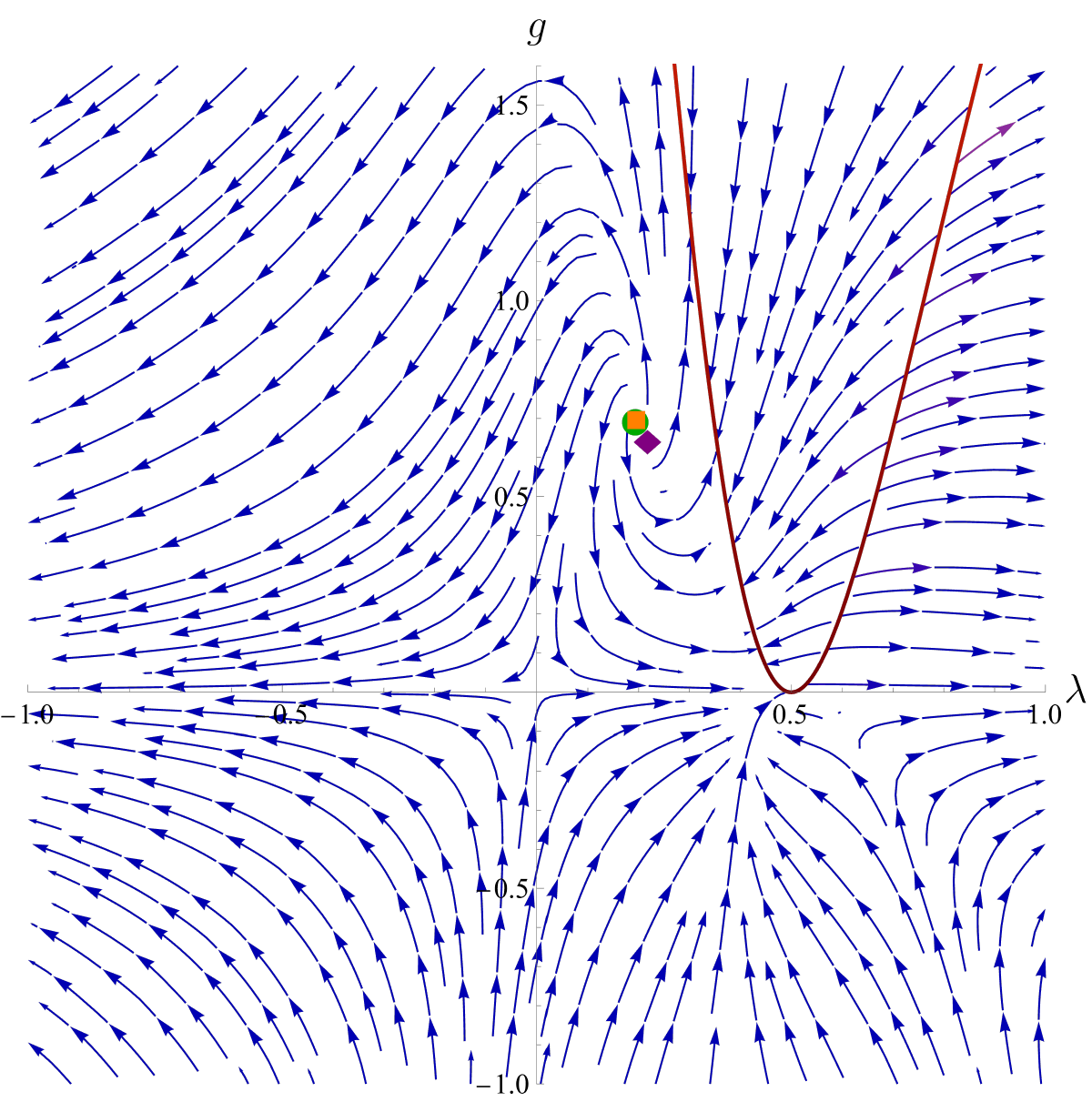}\hfill\includegraphics[width=0.32\textwidth]{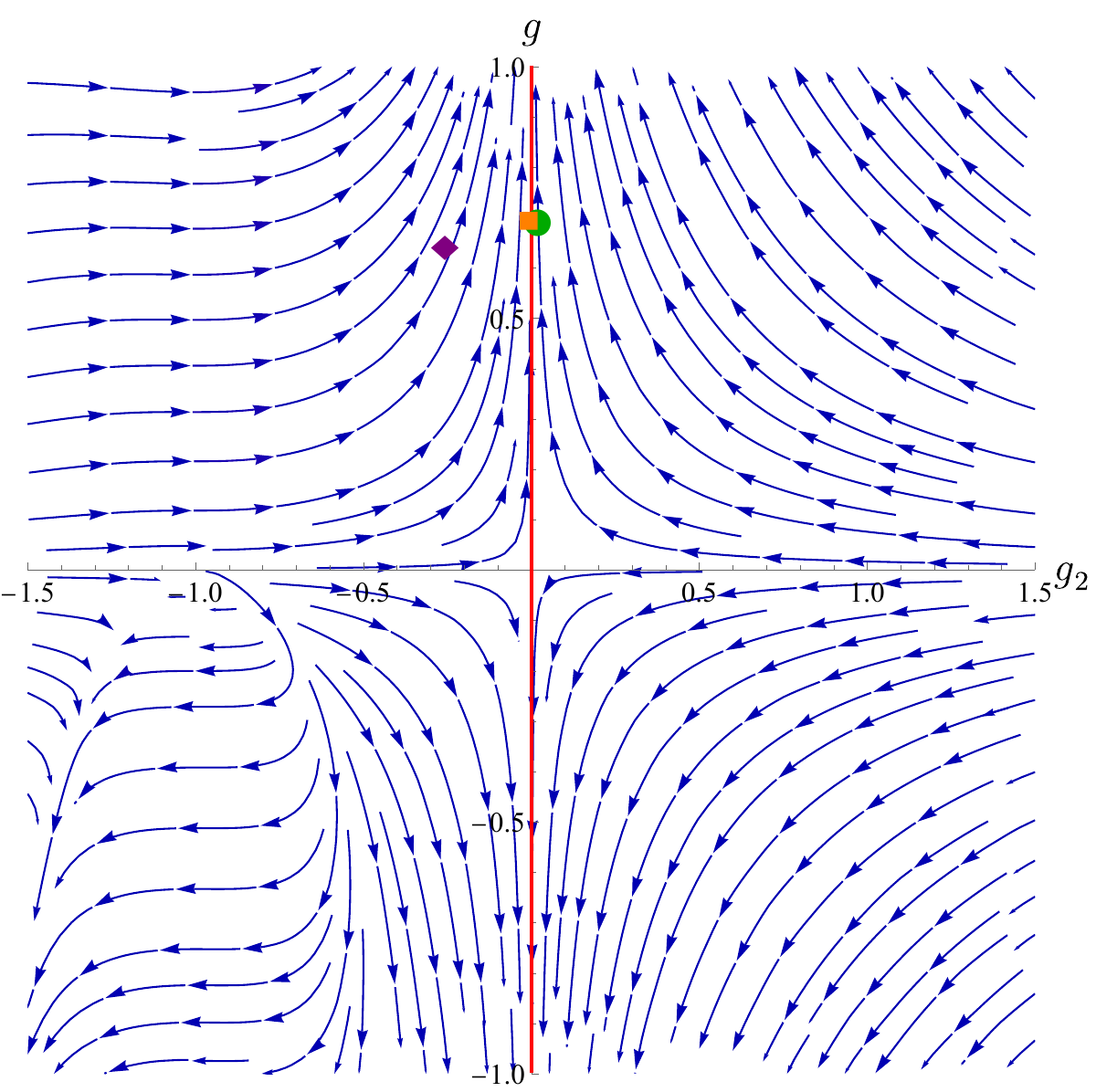}%\hfill\includegraphics[width=0.32\textwidth]{g4a2vsg2Full.pdf}
%\hfill\includegraphics[width=0.32\textwidth]{g4vsg2Full.pdf}
\caption{Projection of the RG flow on the $\{\lambda,g_{4,1}\}$ plane (left panel), $\{\lambda,g\}$ plane (central panel) and $\{g_2,g\}$ plane (right panel). The main, identical-twin, and accidental-twin Proca fixed points are represented by a green circle, an orange square, and a purple rhombus, respectively. In the left and right panels, the red line depicts a line of sGFP---an infinite set of sGFPs, characterized by vanishing couplings, except for $g_{4,1}$ which can take any value. Each sGFP makes the beta functions divergent due to a vanishing Proca mass, but at the same time, such sGFP can act as quasi-fixed points for some RG trajectories. In the central panel, the dark red curve depicts a singular line, where  the beta functions for $g$ and $\lambda$ diverge.} 
\label{fig:g41vsg4Full}
\end{figure*}

In addition to the fixed points described above, there is a GFP, which turns into a fixed-point line starting from the third step of the procedure, i.e., once all operators up to the second order in the Proca field are included. In particular, this is a line of shifted GFPs (sGFPs), since
all couplings vanish except $g_{4,1}$, which is arbitrary.  It is not surprising that the sGFP line is also a singularity of the beta functions: in the limit $g_2\to 0$, the $U(1)$ symmetry is fully restored, rendering the longitudinal mode of the $A_\mu$ field unphysical and causing the Proca propagator to diverge unless a gauge fixing is reintroduced. Nonetheless, the sGFPs can still act as a line of fixed points, since the singularities of the Proca beta functions at $g_2=0$ are of the $1/g_2^n$-type, with $n>0$, and can be eliminated if some of the couplings---particularly, $g(k)$, $g_{4,1}(k)$ and $g_4(k)$---vanish faster than $g_2(k)$. In other words, the sGFP line acts as a line of fixed points for RG trajectories along which $g(k)$, $g_{4,1}(k)$, or $g_4(k)$ vanish more rapidly than $g_2(k)$ as the fixed-point line is approached. Interestingly, also the Reuter fixed point lies on a singular hypersurface of the Proca phase diagram and may be approached as the limit of special RG trajectories. This is unsurprising, as GPTs and standard photon–gravity systems feature different symmetries, and their fixed points are expected to belong to different universality classes. In particular, the standard Reuter fixed point appears at the boundary of validity of the Proca phase diagram, in the limit where the $U(1)$ symmetry is restored. The sGFP line, along with other fixed points and the overall structure of the RG phase diagram can be visualized through projections of the RG flow on sub-theory spaces, cf. Figs.~\ref{fig:3dstep2}-\ref{fig:g41vsg4Full}. 

A final, yet important question is whether the Proca fixed point yields a landscape of effective field theories compatible with positivity bounds and with observations. Since the main Proca fixed point comes with five relevant directions, once the scale of quantum gravity is fixed, the resulting landscape will be parametrized by four dimensionless Wilson coefficient. We reserve this investigation for future work.

\section{Conclusions}

In this work, we have taken first steps toward exploring the UV behavior of GPTs and determining whether these theories admit a UV completion.  

Focusing on a simplified setting with up to two derivatives and four powers in the Proca field, we investigated whether these theories---despite their fine-tuned interaction structure and reduced symmetry---can support a UV fixed point in the spirit of the asymptotic safety program. Our analysis, based on FRG techniques and an appropriate truncation of the effective action, reveals the existence of a triplet of non-trivial fixed points with very similar properties. Among them, only one has a non-tachyonic Proca mass and could thus serve as a consistent UV completion of GPTs. We named this the Proca fixed point. 

Crucially, the properties of the  Proca fixed point differ from those of the well-studied Reuter fixed point of the asymptotic safety scenario for quantum gravity. The Proca fixed point exhibits reduced stability and slower convergence, signaling either that it is a truncation artifact, or that Proca theories have a potentially richer but more delicate RG structure. These features are likely rooted in the non-trivial dynamics defining Generalized Proca models, which is likely not preserved by the RG flow. As such, our findings highlight the necessity of exploring extended truncations. 

Another crucial difference between the theories defined by the Reuter and Proca fixed point lies in their field content: since the Proca field is massive, the $U(1)$ symmetry is broken and the longitudinal mode of the vector field becomes physical. As a result, the RG flow of GPTs is singular in the limit $m\to0$, since in this limit the $U(1)$ symmetry is restored, and a gauge fixing is necessary. In particular, the standard Gaussian and Reuter fixed points lie on singular hypersurfaces of the RG flow of GPTs.

Even if a UV completion for GPTs would be established, their theoretical and phenomenological consistency would still need to be tested. For instance, one would need to investigate whether the landscape (defined along the lines of~\cite{Basile:2021krr,Knorr:2024yiu,DelPorro:2025fiu,DelPorro:2025wts}) resulting from the Proca fixed point is compatible with the positivity bounds in~\cite{deRham:2022sdl,deRham:2018qqo,Bonifacio:2016wcb}. As the Proca fixed point is characterized by five relevant directions, once a unit scale is fixed, the corresponding landscape of effective field theories will be parametrized by four dimensionless Wilson coefficients.

While our results are necessarily preliminary, they provide a proof of principle that UV completions of massive vector field theories, perhaps even beyond GPTs, may be viable. This opens several promising directions for future work, including a systematic classification of fixed points across broader classes of massive vector-tensor theories, the analysis of their phenomenological implications for early universe cosmology, effective field theory, and particle physics, and the investigation of their interplay with quantum gravity.

\begin{acknowledgments}
The authors would like to thank B. Knorr, M. Reichert, and M. Schiffer for discussions, B. Knorr for feedback on the derivation of the RG flow, and G. Lambiase for contributions at early stages of the project. The research of A.P. is supported by a research grant (VIL60819) from VILLUM FONDEN. The Center of Gravity is a Center of Excellence funded by the Danish National Research Foundation under grant No. 184. S.R.A. would like to thank the University of Heidelberg and the University of Copenhagen for hospitality during the development of parts of this work. Moreover, S.R.A. would like to acknowledge the contribution of the COST Action CA23130. This work was supported by the STSM Grant E-COST-GRANT-CA23130-77b4c0df from COST Action CA23130.
\end{acknowledgments}

\vfill

\bibliographystyle{apsrev4-1}
\bibliography{biblio.bib}

\let\clearpage\relax 
\onecolumngrid

\appendix

\section{Propagator expansion: Parker-Fulling expansion}\label{Appendix:PF-exp}

A convenient strategy to handle the right-hand side of the Wetterich equation is to perform a Parker-Fulling expansion of the propagator in powers of derivatives and background fields.

The essential idea is to treat the full inverse propagator as a differential operator, $\mathcal{O} = \mathcal{P} + \mathcal{F} $, where $\mathcal{P}$ is a simple, solvable piece (e.g., the free Klein-Gordon operator in the case of a scalar field theory) and $\mathcal{F}$ contains background-dependent corrections, derivatives of fields, and other interaction terms. One then formally expands the propagator as a geometric series:
\begin{equation}\label{eq:FPexp-simple}
(\mathcal{P} + \mathcal{F})^{-1} = \mathcal{P}^{-1} - \mathcal{P}^{-1} \mathcal{F} \mathcal{P}^{-1} + \mathcal{P}^{-1} \mathcal{F} \mathcal{P}^{-1} \mathcal{F} \mathcal{P}^{-1} + \dots\,,
\end{equation}
and truncates it to an order that matches the truncation of the effective action in the left-hand-side of the Wetterich equation. In particular, each term of the expansion corresponds to a contribution with a definite number of derivatives and powers of the background field(s), which ought to match those appearing in the ansatz for the effective average action.

The inversion of the propagator in GPTs is more involved than in the simple case of Eq.~\eqref{eq:FPexp-simple}. This is because \textit{(i)} the system involves two fields (the Proca and gravitational fields), \textit{(ii)} the two interact non-minimally, and \textit{(iii)} even the individual propagators, $D_i$, are involved. As a result, the inversion of the full regularized propagator, $\mathcal{G}_k=(\Gamma_k^{(2)}+\mathcal{R}_k)^{-1}$, requires requires both block matrix inversion and a double application of the Parker-Fulling expansion. Structurally, the argument of the supertrace in the Wetterich equation reads:
\begin{equation}
    \mathcal{G}_k\cdot \partial \mathbf{R}_k = 
    \begin{pmatrix}
        \text{D}_1 & \text{R} \\
        \text{L} & \text{D}_2
    \end{pmatrix}^{-1}
    \begin{pmatrix}
        \text{I}_1 \,\partial R_k & 0 \\
        0 & \text{I}_2 \,\partial R_k
    \end{pmatrix}
    =\begin{pmatrix}
        (\text{D}_1-\text{R}\text{D}_2^{-1}\text{L})^{-1} & \dots \\
        \dots & (\text{D}_2-\text{L}\text{D}_1^{-1}\text{R})^{-1}
    \end{pmatrix}\begin{pmatrix}
        \text{I}_1 \,\partial R_k & 0 \\
        0 & \text{I}_2 \,\partial R_k
    \end{pmatrix}
\end{equation}
where $\partial R_k\equiv k \partial_k \mathcal{R}_k$ is the RG-time variation of the regulator in the right-hand-side of the Wetterich equation, $R$ and~$L$ are off-diagonal interaction terms, $I_i$ are identity operators in the respective spaces, and the ellipsis denote entries whose specific expressions are unimportant. Each propagator $D_i$ has to be expanded as
\begin{align}
    \text{D}_i^{-1}=&(\text{K}_i+\text{V}_i)^{-1}
    \sim \text{K}_i^{-1}-\text{K}_i^{-1}\text{V}_i\text{K}_i^{-1}+\text{K}_i^{-1}\text{V}_i\text{K}_i^{-1}\text{V}_i\text{K}_i^{-1}\notag\\
    &-\text{K}_i^{-1}\text{V}_i\text{K}_i^{-1}\text{V}_i\text{K}_i^{-1}\text{V}_i\text{K}_i^{-1}+\text{K}_i^{-1}\text{V}_i\text{K}_i^{-1}\text{V}_i\text{K}_i^{-1}\text{V}_i\text{K}_i^{-1}\text{V}_i\text{K}_i^{-1}+\dots\,,
\end{align}
so that the diagonal entries of the block matrix $\mathcal{G}_k$ are
\begin{align}
    (\text{D}_1&-\text{R}\text{D}_2^{-1}\text{L})^{-1} = \text{K}_1^{-1}-\text{K}_1^{-1}\text{V}_1\text{K}_1^{-1}+\text{K}_1^{-1}\text{V}_1\text{K}_1^{-1}\text{V}_1\text{K}_1^{-1}+\text{K}_1^{-1}(\text{R}\text{K}_2^{-1}\text{L})\text{K}_1^{-1}-\text{K}_1^{-1}\text{V}_1\text{K}_1^{-1}\text{V}_1\text{K}_1^{-1}\text{V}_1\text{K}_1^{-1} \nonumber\\
    &-\text{K}_1^{-1}\text{V}_1\text{K}_1^{-1}(\text{R}\text{K}_2^{-1}\text{L})-\text{K}_1^{-1}(\text{R}\text{K}_2^{-1}\text{L})\text{K}_1^{-1}\text{V}_1\text{K}_1^{-1}-\text{K}_1^{-1}(\text{R}\text{K}_2^{-1}\text{V}_2\text{K}_2^{-1}\text{L})\text{K}_1^{-1}\nonumber\\
    &+\text{K}_1^{-1}\text{V}_1\text{K}_1^{-1}\text{V}_1\text{K}_1^{-1}\text{V}_1\text{K}_1^{-1}\text{V}_1\text{K}_1^{-1}+\text{K}_1^{-1}(\text{R}\text{K}_2^{-1}\text{V}_2\text{K}_2^{-1}\text{V}_2\text{K}_2^{-1}\text{L})\text{K}_1^{-1}+\text{K}_1^{-1}(\text{R}\text{K}_2^{-1}\text{L})\text{K}_1^{-1}(\text{R}\text{K}_2^{-1}\text{L})\text{K}_1^{-1}\nonumber\\
    &+\text{K}_1^{-1}\text{V}_1\text{K}_1^{-1}(\text{R}\text{K}_2^{-1}\text{V}_2\text{K}_2^{-1}\text{L})\text{K}_1^{-1}+\text{K}_1^{-1}(\text{R}\text{K}_2^{-1}\text{V}_2\text{K}_2^{-1}\text{L})\text{K}_1^{-1}\text{V}_1\text{K}_1^{-1}+\text{K}_1^{-1}(\text{R}\text{K}_2^{-1}\text{L})\text{K}_1^{-1}\text{V}_1\text{K}_1^{-1}\text{V}_1\text{K}_1^{-1}\nonumber\\
    &+\text{K}_1^{-1}\text{V}_1\text{K}_1^{-1}\text{V}_1\text{K}_1^{-1}(\text{R}\text{K}_2^{-1}\text{L})\text{K}_1^{-1}+\text{K}_1^{-1}\text{V}_1\text{K}_1^{-1}(\text{R}\text{K}_2^{-1}\text{L})\text{K}_1^{-1}\text{V}_1\text{K}_1^{-1}+\mathcal{O}(\text{V}^5)\,,
\end{align} 
and
\begin{align}
    (\text{D}_2&-\text{L}\text{D}_1^{-1}\text{R})^{-1} = \text{K}_2^{-1}-\text{K}_2^{-1}\text{V}_2\text{K}_2^{-1}+\text{K}_2^{-1}\text{V}_2\text{K}_2^{-1}\text{V}_2\text{K}_2^{-1}+\text{K}_2^{-1}(\text{L}\text{K}_1^{-1}\text{R})\text{K}_2^{-1}-\text{K}_2^{-1}\text{V}_2\text{K}_2^{-1}\text{V}_2\text{K}_2^{-1}\text{V}_2\text{K}_2^{-1} \nonumber\\
    &-\text{K}_2^{-1}\text{V}_2\text{K}_2^{-1}(\text{L}\text{K}_1^{-1}\text{R})-\text{K}_2^{-1}(\text{L}\text{K}_1^{-1}\text{R})\text{K}_2^{-1}\text{V}_2\text{K}_2^{-1}-\text{K}_2^{-1}(\text{L}\text{K}_1^{-1}\text{V}_1\text{K}_1^{-1}\text{R})\text{K}_2^{-1}\nonumber\\
    &+\text{K}_2^{-1}\text{V}_2\text{K}_2^{-1}\text{V}_2\text{K}_2^{-1}\text{V}_2\text{K}_2^{-1}\text{V}_2\text{K}_2^{-1}+\text{K}_2^{-1}(\text{L}\text{K}_1^{-1}\text{V}_1\text{K}_1^{-1}\text{V}_1\text{K}_1^{-1}\text{R})\text{K}_2^{-1}+\text{K}_2^{-1}(\text{L}\text{K}_1^{-1}\text{R})\text{K}_2^{-1}(\text{L}\text{K}_1^{-1}\text{R})\text{K}_2^{-1}\nonumber\\
    &+\text{K}_2^{-1}\text{V}_2\text{K}_2^{-1}(\text{L}\text{K}_1^{-1}\text{V}_1\text{K}_1^{-1}\text{R})\text{K}_2^{-1}+\text{K}_2^{-1}(\text{L}\text{K}_1^{-1}\text{V}_1\text{K}_1^{-1}\text{R})\text{K}_2^{-1}\text{V}_2\text{K}_2^{-1}+\text{K}_2^{-1}(\text{L}\text{K}_1^{-1}\text{R})\text{K}_2^{-1}\text{V}_2\text{K}_2^{-1}\text{V}_2\text{K}_2^{-1}\nonumber\\
    &+\text{K}_2^{-1}\text{V}_2\text{K}_2^{-1}\text{V}_2\text{K}_2^{-1}(\text{L}\text{K}_1^{-1}\text{R})\text{K}_2^{-1}+\text{K}_2^{-1}\text{V}_2\text{K}_2^{-1}(\text{L}\text{K}_1^{-1}\text{R})\text{K}_2^{-1}\text{V}_2\text{K}_2^{-1}+\mathcal{O}(\text{V}^5)\,,
\end{align}
where the truncation to the order $\mathcal{O}(\text{V}^5)$ is chosen to match our ansatz~\eqref{FullL}, which contains up to two derivatives and up to four powers of the Proca field.

Each of the terms in the sums above was computed using \texttt{xact}~\cite{Martin-Garcia:2007bqa}, and contains curvature invariants, open derivatives, and functions of the Laplace operators. These must be appropriately rearranged in order to apply standard heat-kernel techniques. This requires the use of commutation relations, which we summarize in the next appendix.

\section{Commutator rules}\label{Appendix:Commutator_rules}

Computing the right-hand-side of the Wetterich equation, Eq.~\eqref{Wetterich_eq}, requires evaluating the supertrace of the regularized propagator, contracted with the $k$-variation of the regulator $\mathcal{R}_k$. As mentioned in the main text, the dynamics~\eqref{FullL} generates a plethora of terms involving functions of the Laplacian operator, open derivatives, and tensors. Prior to evaluating the trace, we need to appropriately re-order these operators.
 The main idea is to write each term appearing in the trace of the Wetterich equation such that Laplacian functions are in between tensors, which should be moved to the left, and open derivatives, which ought to be on the right. In this way, the usual heat kernel techniques can be directly applied. The re-ordering of operators requires accounting for the commutators between arbitrary functions of the Laplacian, tensors and open derivatives. If $f$ is a suitable function of the (background) Laplacian $\bar{\Delta}$, $W$ is an arbitrary operator and $X$ a tensor of arbitrary rank, then~\cite{Knorr:2021slg} 
\begin{equation}
    f(\bar{\Delta})WX=Wf(\bar{\Delta})X+\sum_{m\geq 1}\frac{1}{m!}\sum_{n=0}^{m-1}\left(-1\right)^{m-1-n}\binom{m-1}{n}\bar{\Delta}^n\left[\bar{\Delta},W\right]\bar{\Delta}^{m-n-1}f^{(m)} (\bar{\Delta})X \,.
    \label{App_CommRules_Generic}
\end{equation}
In our case, $W$ is a ``multiplication operator'' in the sense of~\cite{Knorr:2021slg}, and hence
\begin{equation}
    \left[ \bar{\Delta},W \right] = \left(\bar{\Delta}W \right)-2\left(\bar{\nabla}^\mu W\right)\bar{\nabla}_\mu \, .
\end{equation}
Replacing this commutator in Eq.~\eqref{App_CommRules_Generic}, one gets~\cite{Knorr:2021slg}
\begin{equation}
    f(\bar{\Delta})WX=W\,f(\bar{\Delta})X+\sum_{m\geq 1}\frac{1}{m!}\sum_{n=0}^{m-1}\left(-1\right)^{m-1-n}\binom{m-1}{n}\bar{\Delta}^n\left\{ \left(\bar{\Delta}W \right)-2\left(\bar{\Delta}^\mu W\right)\bar{\nabla}_\mu\right\}\bar{\Delta}^{m-n-1}f^{(m)} (\bar{\Delta})X \, .
\end{equation}
These commutation rules must be iteratively applied in the code, to each term in the argument of the supertrace, until the right ordering of operators is reached. At each iteration one can simplify expressions by neglecting operators beyond our truncation, Eq.~\eqref{FullL}. For instance, each commutator increases the derivative order by at least one. Since our truncation includes terms up to second order in derivatives, the series above must be consistently truncated so that only contributions up to this order are retained.

\section{Fixed points and critical exponents}\label{appendix:Tables}

This appendix collects the numerical results underlying the discussion in Sect.~\ref{Sec_Asymptotic_safety_in_Generalized_Proca_Theories}.
For each truncation order considered in our analysis, which we increase in a step-by-step fashion, we list the fixed-point values of the dimensionless couplings and the corresponding critical exponents.
Starting from the full-$A^2$ truncation, where the Proca fixed point first appears, the ordering of the fixed points follows the convention introduced in the main text: the main Proca fixed point is reported in the first row of each table, its identical twin in the second row, and the accidental twin in the third row. The sGFP line is omitted in the tables. 

The lowest-order truncation that one can consider in our system is the EH truncation,
\begin{equation}\label{eq:tr1}
    \mathcal{L}_{EH} = \frac{1}{16\pi G}\left(2\Lambda-R\right) \,.
\end{equation}
In this first step one finds the Reuter fixed point of the asymptotic safety program, with fixed-point values and corresponding critical exponents (cf. Tab.~\ref{tab:FP&CritcExpEH}) in agreement with the results in literature~\cite{Reuter:2001ag}. 

\begin{table}[t!]
    \centering
    \begin{tabular}{|c|c|c|c|}
        \hline
        \multicolumn{4}{|c|}{Einstein-Hilbert truncation }\\
        \hline
        \noalign{\vskip -1pt}
        \hline
        \multicolumn{2}{|c|}{Fixed Point} & \multicolumn{2}{c|}{Critical Exponents} \\ 
        \hline
        $\lambda^*$ & $g^*$ & $\theta_1$ & $\theta_2$ \\ 
        \hline
        $0.1969$ & $0.6907$ & $2.380 + i\,2.821 $ & $2.380- i\,2.821$ \\ 
        \hline
    \end{tabular}
    \caption{NGFP and respective critical exponents in the EH truncation. This coincides with the Reuter fixed point of the asymptotic safety program.}
    \label{tab:FP&CritcExpEH}
\end{table}
The second step is the first non-trivial truncation of GPTs, since a massive vector field minimally coupled to gravity is now included,
\begin{equation}\label{eq:tr2}
    \mathcal{L}_{EH+G_2} = \frac{1}{16\pi G}\left(2\Lambda-R\right)+\frac{1}{4}F_{\mu\nu}F^{\mu\nu}+\frac{1}{2}G_2 A^2\,.
\end{equation}
In this truncation there are three dimensionless couplings, i.e., $\{\lambda,g,g_2\}$. As already detailed in the main text, the running of the coupling $Z_A$, as well as other derivative couplings, cannot be resolved due to our choice of background. As shown in Tab.~\ref{tab:FP&CritExpEHG2} and as detailed in the main text, in this truncation GPTs admit three fixed points. The first fixed point is a massless NGFP, which can be identified again with the Reuter fixed point and is the continuation of the fixed point found in the first step. The other two fixed points, listed in Tab.~\ref{tab:FP&CritExpEHG2} for completeness, are instead spurious, since they disappear (or, they become and stay complex) in higher-order truncations. 

\begin{table}[t!]
\centering
\begin{tabular}{|c|c|c|c|c|c|}
    \hline
    \multicolumn{6}{|c|}{Proca field minimally coupled to gravity}\\
        \hline
        \noalign{\vskip -1.0pt}
        \hline
    \multicolumn{3}{|c|}{Fixed Points}&\multicolumn{3}{c|}{Critical Exponents} \\ 
    \hline
    $\lambda^*$ & $g^*$ & $g_2^*$ & $\theta_1$ & $\theta_2$ & $\theta_3$  \\
    \hline
    $0.1969$ & $0.6907$ & $0$ & $2.380 + i\, 2.821$ & $2.380 - i\, 2.821 $ & $1.202$\\
    \hline
    $0.1721$ & $0.7612$ & $0.7652$ & $2.120 + i\, 2.573$ & $2.120 - i\, 2.573 $ & $-0.5014$\\
    \hline
    $0.04167$ & $1.048$ & $-2.595$ &$3.163$ & $-1.769$ & $1.788$\\
    \hline
    \end{tabular}
    \caption{NGFPs and respective critical exponents for a Proca field minimally coupled to gravity, Eq.~\eqref{eq:tr2}. Once the Proca field is added, the flow is modified so to admit three fixed points, and at this stage any of them could in principle be a suitable candidate for the UV completion of GPTs. In the table, the first line is a massless NGFP, again identified with the Reuter fixed point of the asymptotic safety program. The other two fixed points turn out to be spurious, as this is the only truncation in which they appear as real fixed points.}
        \label{tab:FP&CritExpEHG2}
    \end{table}
    
Next, we included the lowest-order non-minimal interaction between the Proca and gravitational field. The resulting dynamics is a truncation of GPTs including all operators up to quadratic order in the Proca field,
\begin{equation}\label{eq:tr3}
    \mathcal{L}_{\text{Full}\,A^2} = \frac{1}{16\pi G}\left(2\Lambda-R\right)+\frac{1}{4}F_{\mu\nu}F^{\mu\nu}+\frac{1}{2}G_2\,A^2 + G_{4,1}\left[A^2\,R + \left(\nabla_\mu A^\mu\right)^2 - \left(\nabla_\mu A_\nu\right)\left(\nabla^\nu A^\mu\right)\right]\,.
\end{equation} 
Fixed points and critical exponents for this case are reported in Tab.~\ref{tab:FP&CritExpA2}. By including the coupling $g_{4,1}$ into the system, the fixed-point structure becomes reacher, it being governed by three fixed points with similar fixed-point values: the main Proca fixed point, which appears as a complex-conjugate pair at this stage, and two ``twin'' Proca fixed points with tachyonic mass. These fixed points are not generated by a fixed point collision, as shown in Fig.~\ref{fig:quasi}. The discontinuity in the fixed-point structure also prevents us from tracking the properties of individual fixed points when moving from the second to the third step of our analysis.

\begin{table}[t!]
    \centering
    \begin{tabular}{|c|c|c|c|}
        \hline
        \multicolumn{4}{|c|}{Full $A^2$ truncation}\\
        \hline
        \noalign{\vskip -1.0pt}
        \hline
        \multicolumn{4}{|c|}{Fixed Points}\\ 
        \hline
        $\lambda^*$ & $g^*$ & $g_2^*$ & $g_{4,1}^*$ \\ 
        \hline
        $0.1942\pm i\,0.0004670 $ & $0.6953\pm i\,0.0006311$ & $0.03232\pm i\,0.004362$ & $0.01872\mp i\,0.01085$ \\
        \hline
        $0.2076$ & $0.6790$ & $-0.07270$ & $-0.09667$\\ 
        \hline
        $0.2824$ & $0.5202$ &  $-0.5899$ & $-0.1883$\\
        \hline
        \multicolumn{4}{|c|}{Critical Exponents} \\
        \hline
        $\theta_1$ & $\theta_2$ & $\theta_3$ & $\theta_4$ \\
        \hline
        $2.332 \pm i\,2.807$ & $2.381\mp i\,2.794$ & $1.810\pm i\,5.642$ & $1.066 \mp i\,0.1330$ \\
        \hline
        $2.497 + i\,2.987$ &  $2.497 - i\,2.987$  & $-1.787$ & $-16.77$ \\
        \hline
        $5.476 + i\, 4.457 $ & $5.476 - i\, 4.457 $ & $23.83 $ & $-14.98$\\
        \hline
    \end{tabular}
    \caption{NGFPs and critical exponents for GPTs truncated to second order in a field expansion, Eq.~\eqref{eq:tr3}. In this truncation, the pure-EH fixed point of the previous step has split into three qualitatively new NGFPs---the Proca triplet, whereas the two spurious fixed points have disappeared. The main Proca fixed point (first line) emerges at this step as a pair of complex-conjugate fixed points, suggesting that this candidate UV completion of GPTs may be spurious or unstable. The other two twin Proca fixed points are real, but describe a tachyonic Proca field.}
    \label{tab:FP&CritExpA2}
\end{table}

Adding another coupling to the truncation, specifically including the $G_4 A^4$ operator, the truncated Lagrangian reads
\begin{equation}\label{eq:tr4}
    \mathcal{L}_{\text{Full}\,A^2+G_4} = \frac{1}{16\pi G}\left(2\Lambda-R\right)+\frac{1}{4}F_{\mu\nu}F^{\mu\nu}+\frac{1}{2}G_2\,A^2 + G_{4,1}\left[A^2\,R + \left(\nabla_\mu A^\mu\right)^2 - \left(\nabla_\mu A_\nu\right)\left(\nabla^\nu A^\mu\right)\right] + G_4A^4\,.
\end{equation}
The corresponding fixed points are listed in Tab.~\ref{tab:FP&CritExpA2+A4}. The main Proca fixed point turns real at this step, making it a viable candidate for the UV completion of GPTs. The $g_2^\ast$ coupling associated with both twin Proca fixed points is negative, making once again the Proca field emerging from these fixed points tachyonic. The critical exponents evolve consistently from the previous to this step, but they are comparatively less stable than those of the Reuter fixed point~\cite{Falls:2014tra}.

\begin{table}[t!]
    \centering
    \begin{tabular}{|c|c|c|c|c|}
        \hline
        \multicolumn{5}{|c|}{Full $A^2$ truncation, complemented by the $G_4$ coupling}\\
        \hline
        \noalign{\vskip -1.0pt}
        \hline
        \multicolumn{5}{|c|}{Fixed Points} \\ 
        \hline
        $\lambda^*$ & $g^*$ & $g_2^*$ & $g_{4,1}^*$ & $g_4^*$ \\
        \hline
        $0.1928$ & $0.6983$ & $0.05517$ & $0.02531$ & $-0.2439$ \\
        \hline
        $0.1969$ & $0.6897$ & $-0.006966$ & $0.005455$ & $0.3060$ \\ 
        \hline
        $0.2507$ & $0.5806$  &$-0.4670$ & $-0.1298$ & $1.109$ \\
        \hline
        \multicolumn{5}{|c|}{Critical Exponents}
        \\
        \hline
        $\theta_1$ & $\theta_2$ & $\theta_3$ & $\theta_4$ & $\theta_5$\\
        \hline
        $2.368 + i\, 2.770$ & $2.368 - i\, 2.770$ & $-0.6771 +i\,1.028$ & $-0.6771 - i\,1.028$ & $1.931$\\
        \hline
        $2.377 + i\, 2.817 $ & $2.377 - i\, 2.817 $ & $2.271 + i\, 7.807 $ & $2.271 - i\, 7.807 $ &  $4.659$\\
        \hline
        $3.408 +i\, 3.850$ & $3.408  - i\,3.850 $ & $19.35$ & $-8.409$ & $17.34$  \\
        \hline
    \end{tabular}
    \caption{NGFPs and critical exponents for  GPTs, including operators up to second order in the Proca field and one additional $A^4-$term, Eq.~\eqref{eq:tr4}. The addition of the $A^4-$coupling, $g_4$, stabilizes the main Proca fixed point, making it real. The two twin Proca fixed points remain real, but their Proca mass is still tachyonic, rendering them physically inconsistent.}
    \label{tab:FP&CritExpA2+A4}
\end{table}

\newpage In the last step, all operators of Eq.~\eqref{FullL} are taken into account. Fixed points and critical exponents are given in Tab.~\ref{tab:FP&CritExpFull}. We note that the identical-twin Proca fixed point is the most predictive within the Proca triplet, as it features four relevant directions. The critical exponents of the accidental-twin Proca fixed point are somewhat stable across truncations but their magnitude indicate a highly non-perturbative nature. Both twin Proca fixed points remain tachyonic, making the main Proca fixed point the sole viable candidate for a consistent UV completion of GPTs. 

\begin{figure*}[t!]
\includegraphics[width=0.50\textwidth]{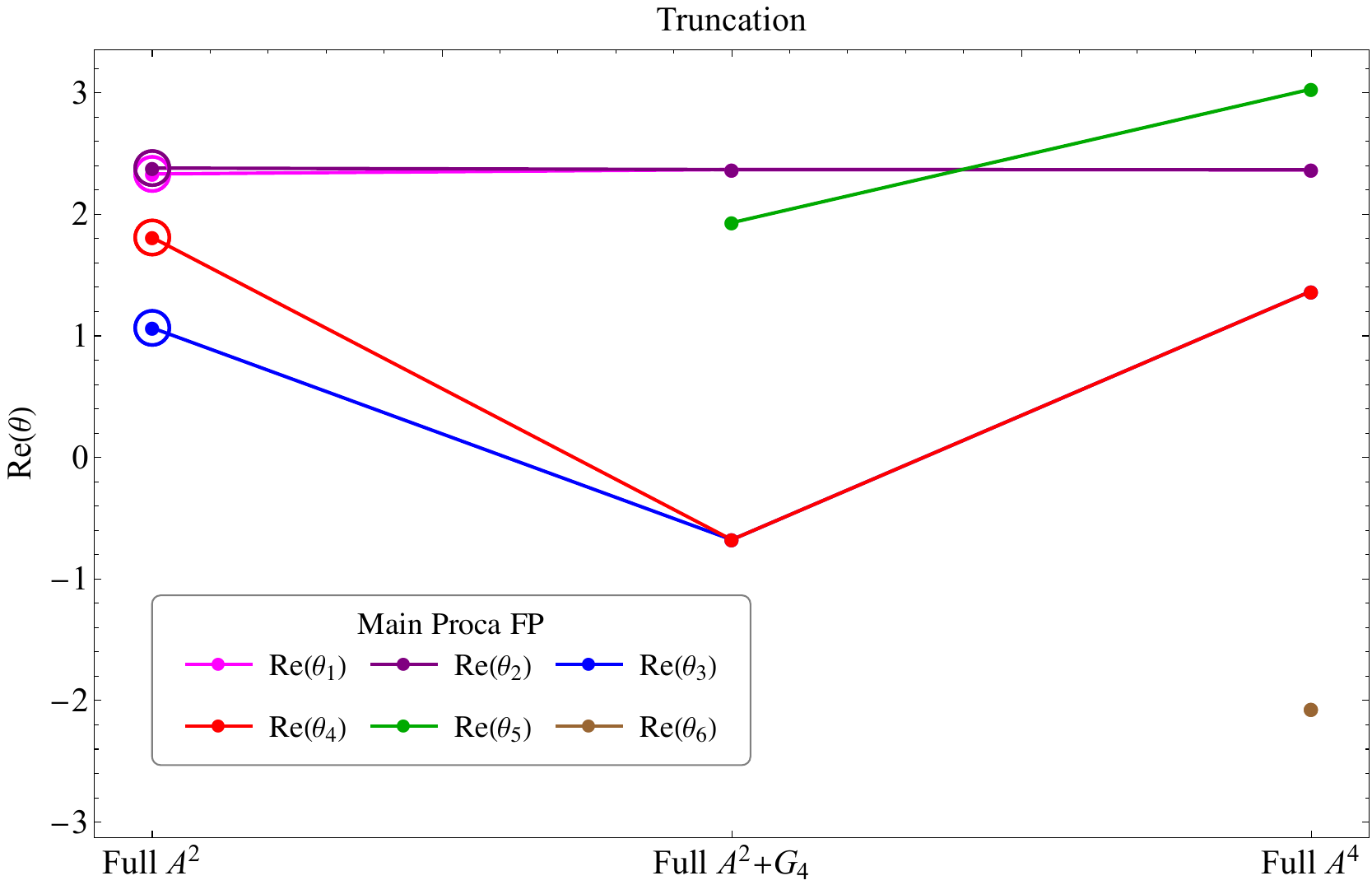}\hfill\includegraphics[width=0.50\textwidth]{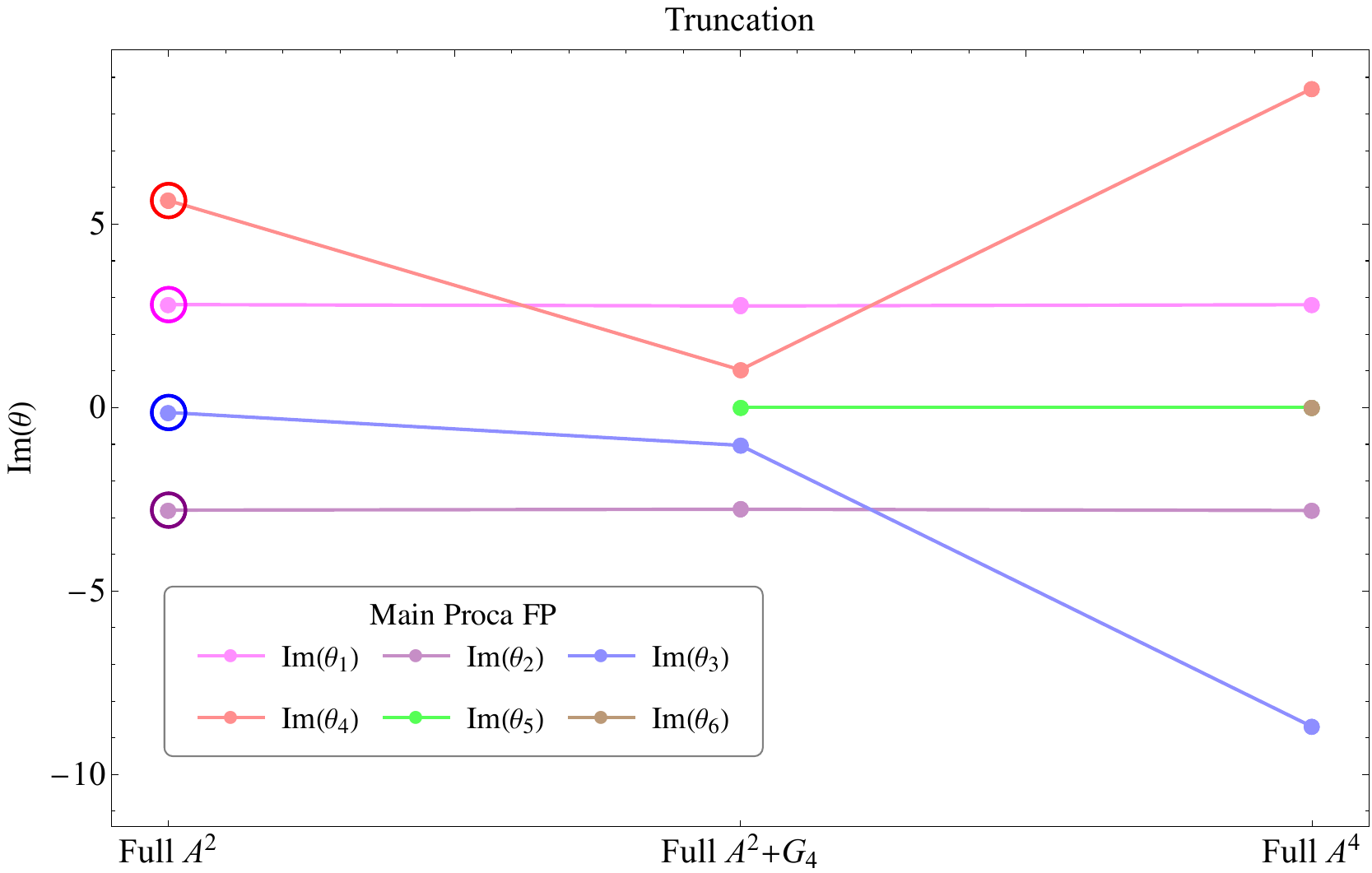}\hfill\includegraphics[width=0.50\textwidth]{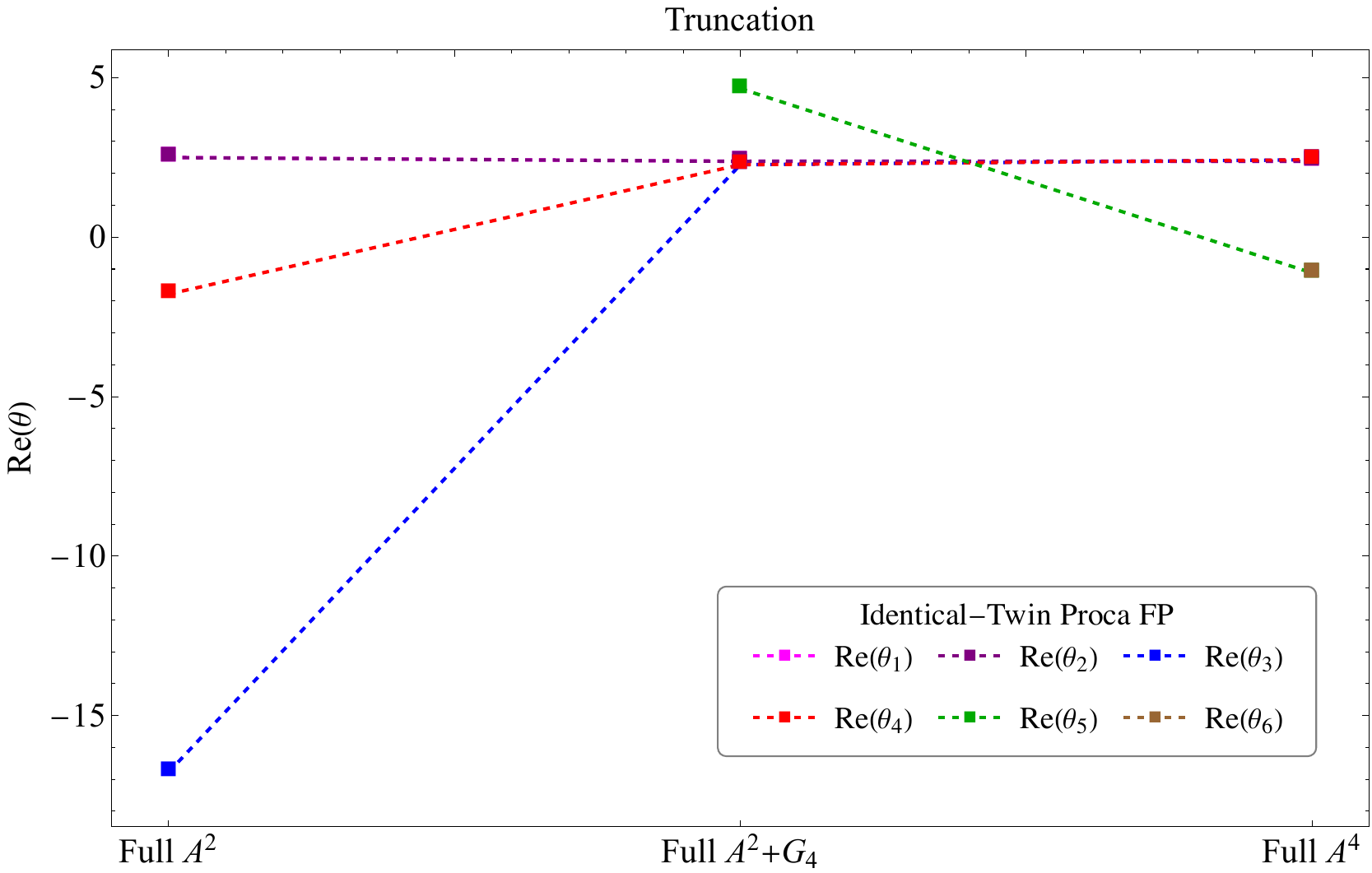}\hfill\includegraphics[width=0.50\textwidth]{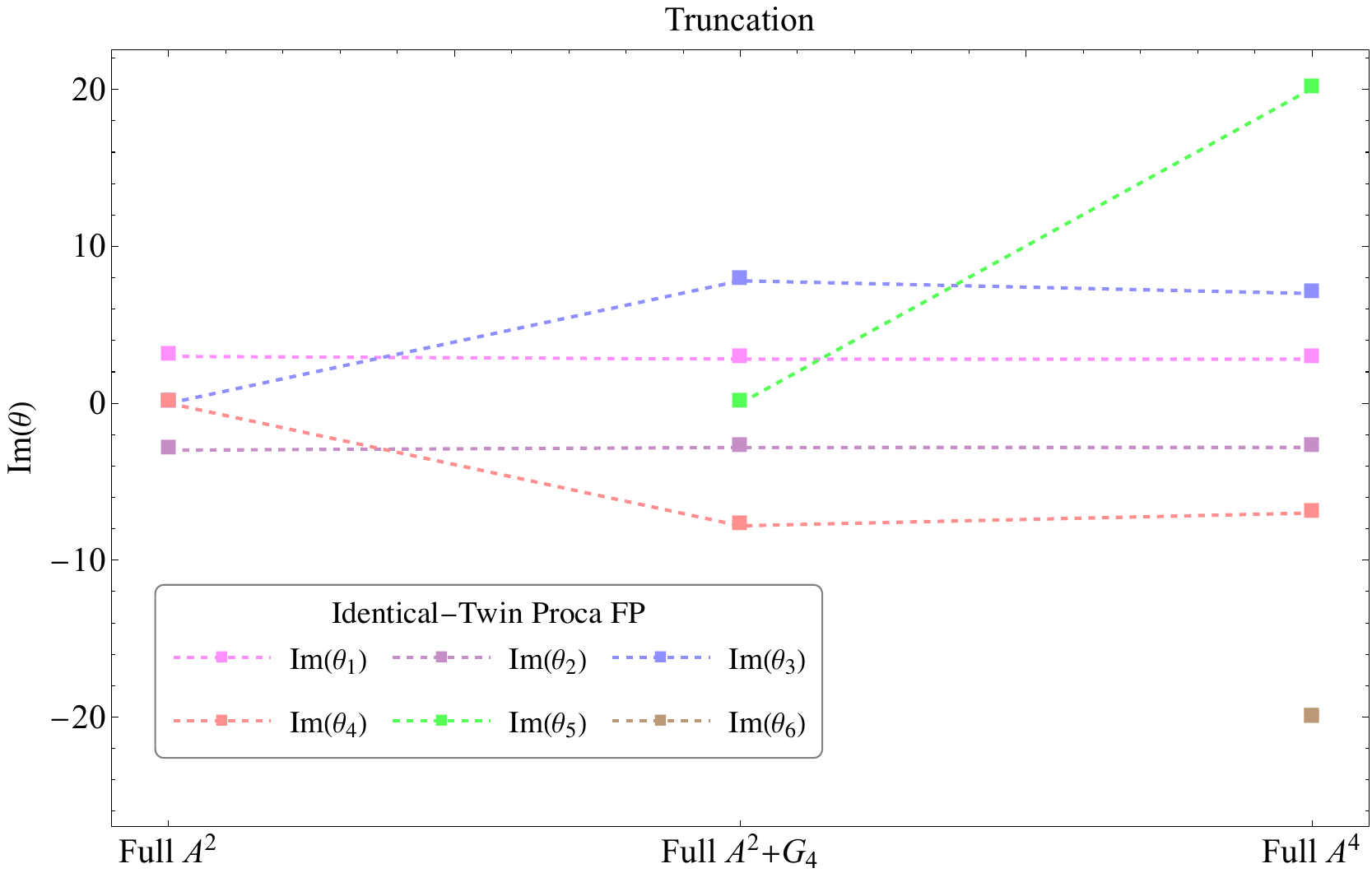}\hfill\includegraphics[width=0.50\textwidth]{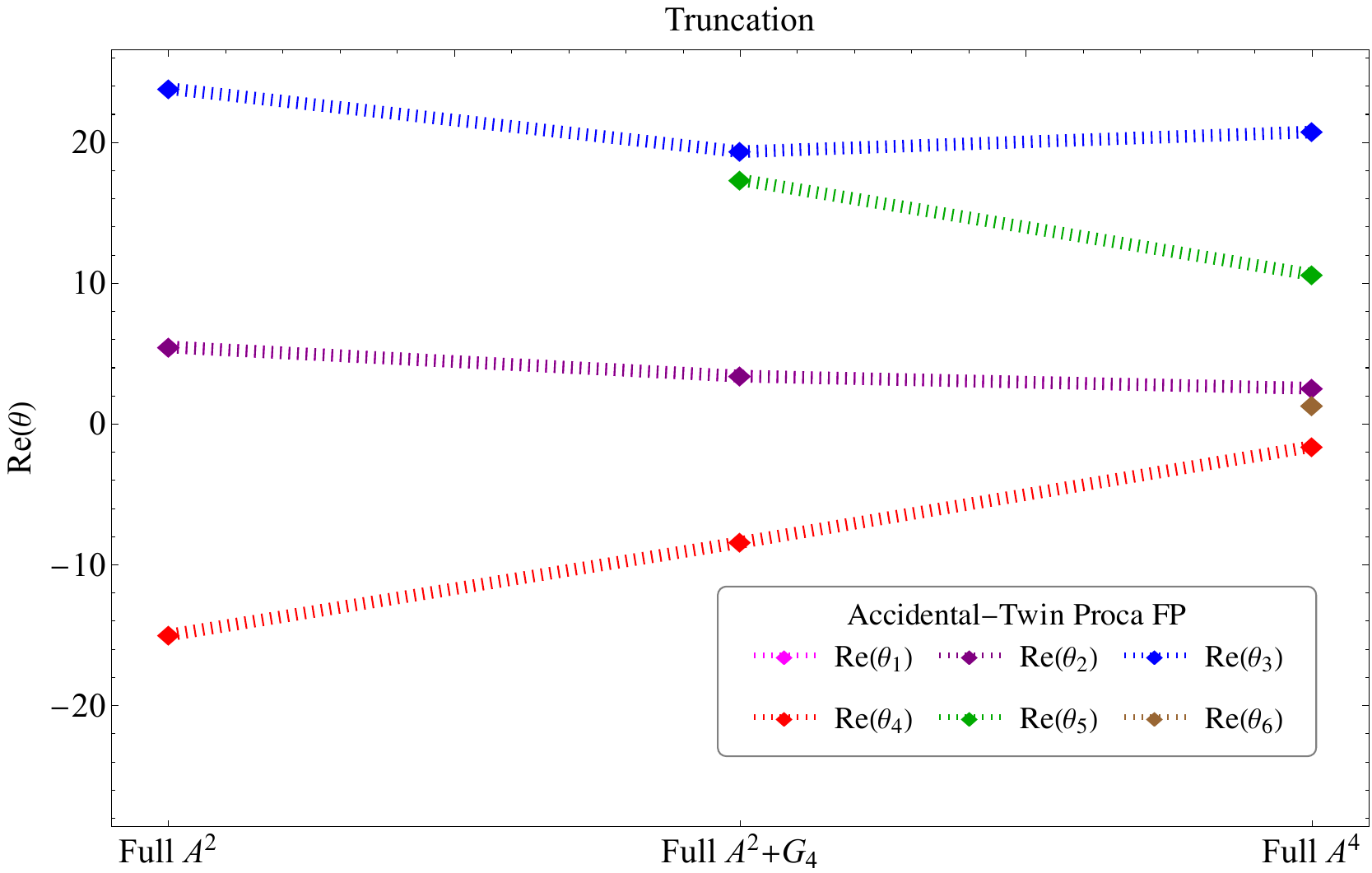}\hfill\includegraphics[width=0.50\textwidth]{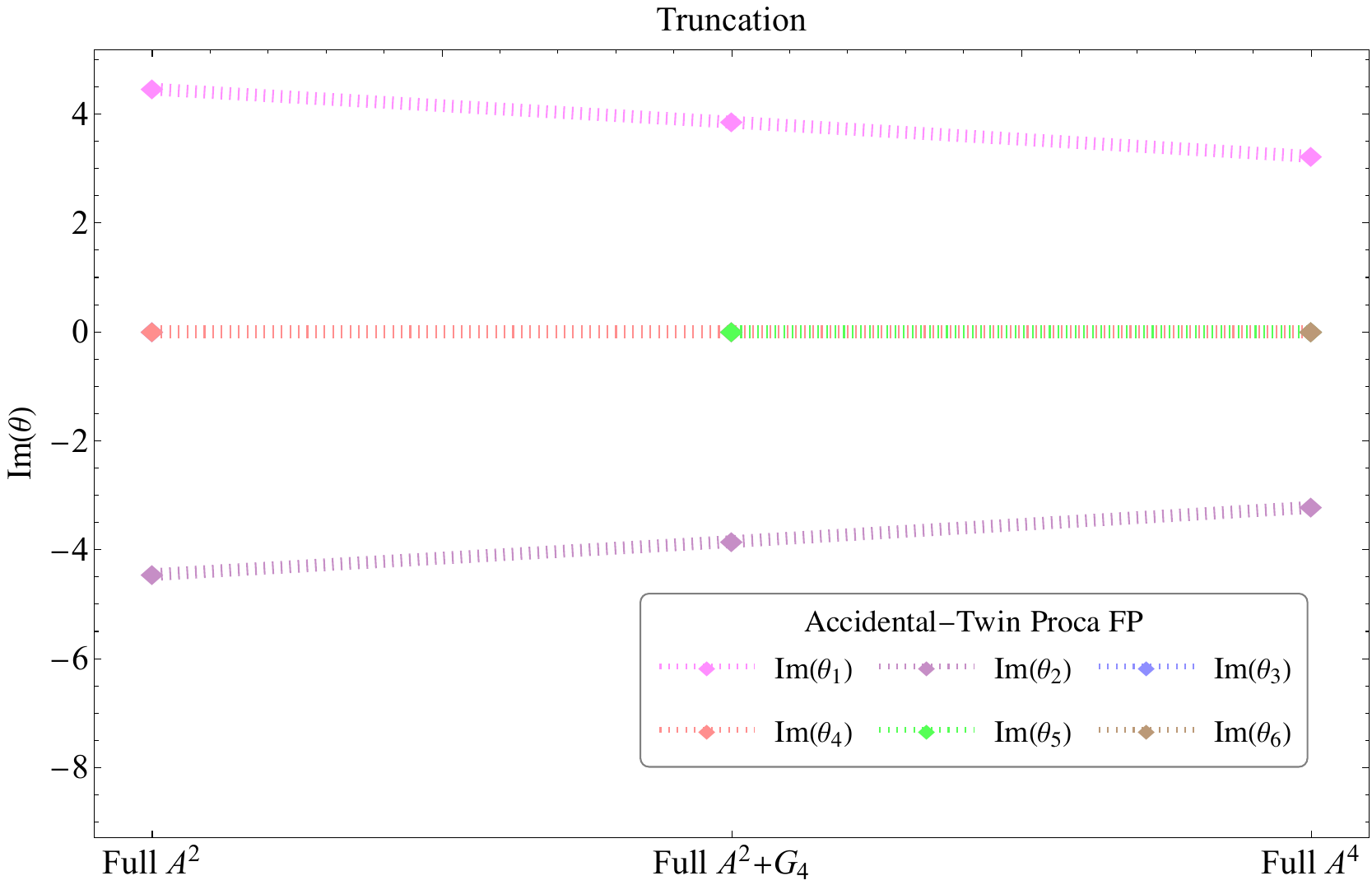}
\caption{Evolution of the real $\text{Re}(\theta)$ (left panels) and imaginary $\text{Im}(\theta)$ (right panels) parts of the critical exponents of the main (top panels), identical-twin (central panels) and accidental-twin (bottom panels) Proca fixed points as a function of the truncation order. The associated fixed-point values evolve with the truncation according to Fig.~\ref{fig:FPvsTruncation}. The convention for markers and lines is as in Fig.~\ref{fig:FPvsTruncation}. In particular, filled markers correspond to the critical exponents of real fixed points, while empty markers stand for those of complex-conjugate fixed points. Moreover, lighter colors for lines in the right panels depict the evolution of the imaginary parts of the critical exponents. For all fixed points of the Proca triplet, the critical exponents associated with the first two eigendirections are quite stable, while the other ones vary substantially, indicating either that the fixed points are spurious or that their convergence is slow.}
    \label{fig:ReCritExpvsTruncation}
\end{figure*}

The evolution of the critical exponents across increasing truncations is summarized in Fig.~\ref{fig:ReCritExpvsTruncation} for the three fixed points of the Proca triplet. For each fixed point, we display separately the real and imaginary parts of the critical exponents as functions of the truncation order. From the six figures, it becomes visually evident that the critical exponents associated with the first two eigendirections exhibit a high degree of stability under changes in the truncation. By contrast, the remaining critical exponents display a stronger dependence on the truncation order, with significant variations in both their real and imaginary parts.

Overall, based on the evolution of its fixed-point coordinates and critical exponents---most notably when compared with those of the Reuter fixed point of the asymptotic safety program---the Proca triplets seem to converge slowly. Hence, assessing the existence and physical relevance of the Proca triplet---and in particular of the main Proca fixed point as a candidate for a consistent UV completion of GPTs---requires extending the truncation beyond the present level. Higher-order  operators may either stabilize the critical exponents and provide more evidence for the existence of the Proca triplet, or lead to the disappearance of some of its fixed points.

\begin{table}[t!]
    \centering
    \begin{tabular}{|c|c|c|c|c|c|}
        \hline
        \multicolumn{6}{|c|}{Full $A^4$ truncation}\\
        \hline
        \noalign{\vskip -1.0pt}
        \hline
        \multicolumn{6}{|c|}{Fixed Points} \\ 
        \hline
        $\lambda^*$ & $g^*$ & $g_2^*$ & $g_{4,1}^*$ & $g_4^*$ & $g_{4,2}^*$ \\ 
        \hline
        $0.1955$ & $0.6937$ & $0.02084$ & $0.005806$ & $-0.1050$ & $0.9466$\\
        \hline
        $0.1968 $ & $0.6897$ & $-0.007474$ & $0.008201$ & $0.4806$ & $-0.3028$ \\ 
        \hline
        $ 0.2185$ & $0.6440$ & $-0.2571$ & $-0.06036 $ & $1.201 $ & $1.190 $ \\
        \hline
        \multicolumn{6}{|c|}{Critical Exponents}
        \\
        \hline
        $\theta_1$ & $\theta_2$ & $\theta_3$ & $\theta_4$ & $\theta_5$ & $\theta_6$\\
        \hline
        $2.367 + i\, 2.804 $ & $2.367 - i\, 2.804 $ & $1.368 + i\, 8.694$ & $1.368 - i\,8.694$  & $ 3.030 $ & $-2.068$\\
        \hline
        $2.380 + i\, 2.813$ & $2.380  - i\, 2.813$ & $2.426 + i\, 6.994 $ & $2.426 - i\, 6.994 $ & $-1.120 +i\, 20.06 $ & $-1.120 -\, i\,20.06 $  \\
        \hline
        $2.545 + i\, 3.222 $ & $2.545  -i\, 3.222$ & $20.75$ & $-1.593$  & $10.62 $  & $1.321$ \\
        \hline
    \end{tabular}
    \caption{NGFP and critical exponents for the full truncation, Eq.~\eqref{FullL}. The Proca fixed point is the only fixed point with non-tachyonic mass, making it the only consistent candidate UV completion of GPTs. It comes with five relevant directions, meaning that the  set of effective field theories it leads to is parametrized by four dimensionless parameters. Yet, it is important to remark that it exhibits slow convergence: while its fixed-point are mostly stable, many of the corresponding critical exponents vary substantially across truncations.}
    \label{tab:FP&CritExpFull}
\end{table}

\vfill

\end{document}